\documentclass{aa}

\usepackage{graphicx}
\usepackage{float}
\usepackage{chngpage}
\usepackage{calc}
\usepackage[colorlinks=true,linkcolor=black, citecolor=blue,urlcolor=blue]{hyperref}
\begin{document} 
  
  \title{Inferring the star formation histories of the most massive and passive early-type galaxies at z<0.3}
  

   \author{Annalisa Citro,
          \inst{1,2}
          Lucia Pozzetti,
          \inst{2} Michele Moresco,\inst{1,2} Andrea Cimatti \inst{1}}

  \institute{
              Dipartimento di Fisica e Astronomia, Universit\`{a} di Bologna, Viale Berti Pichat 6/2, I-40127 Bologna, Italy\\
              \email{annalisa.citro@unibo.it
             }
         \and
             INAF $-$Osservatorio Astronomico di Bologna, via Ranzani 1, 40127 Bologna, Italy\\
             }
	\titlerunning{Inferring the star formation histories of the most massive and passive early-type galaxies at z<0.3}
		\authorrunning{A.Citro et al. }

  \date{...}
 
 \abstract
 {
In the $\Lambda$CDM cosmological framework, massive galaxies are the end-points of the 
hierarchical evolution and are therefore key probes to understand how the baryonic matter 
evolves within the dark matter halos. 
}
 {
The aim of this work is to use the "archaeological" approach in order to infer the stellar 
population properties and star formation histories of the most massive ($M>10^{10.75}M_{\sun}$) and passive early-type 
galaxies (ETGs) at $0<z<0.3$ (corresponding to a cosmic time interval of $\sim3.3$ Gyr) based on stacked, high signal-to-noise ratio (SNR), spectra extracted
from the Sloan Digital Sky Survey (SDSS). Our study is focused on the most passive 
ETGs in order to avoid the contamination of galaxies with residual star formation activity and
extract the evolutionary information on the oldest envelope of the global galaxy population.

}
 {
Differently from most previous studies in this field, we did not rely on individual absorption
features such as the Lick indices, but we exploited the information present in the full 
spectrum by means of the STARLIGHT public code, adopting different stellar population synthesis models.
Successful tests have been performed in order to assess the reliability of STARLIGHT to
retrieve the evolutionary properties of the ETG stellar populations such as age, 
metallicity and star formation history. The results indicate that these properties can be
derived with a percentage accuracy better than $10~\%$ at SNR$~\gtrsim~$10 -- 20, and also that the procedure of stacking galaxy spectra does not introduce significant biases on their retrieving.
}
 {
Based on our spectral analysis, we found that the ETGs of our sample are very old systems
with ages almost as old as the age of the Universe in the case of the most massive ones.
The stellar metallicities are slightly supersolar, with a mean of $Z\sim0.027\pm0.002$ and $Z\sim0.029\pm0.0015$ (depending on the spectral synthesis models used for the fit) and do not depend on 
redshift. Dust extinction is very low, with a mean of $A_{V}\sim0.08\pm0.030~$mag and $A_{V}\sim0.16\pm0.048~$mag. The ETGs show an anti-hierarchical 
evolution (downsizing) where more massive galaxies are older. 
The SFHs can be approximated with a parametric function of the form $SFR(t)\propto \tau^{-(c+1)}t^{c}~exp(-t/\tau)$ , with typical short $e$-folding times of $\tau\sim0.6-0.8$ Gyr  (with a dispersion of $\pm~0.1$ Gyr) and $c\sim0.1$ (with a dispersion of $\pm~0.05$).
Based on the reconstructed
SFHs, most of the stellar mass ($\gtrsim75~\%$) was assembled by $z\sim5$ and $\lesssim4~ \%$ 
of it can be ascribed to stellar populations younger than $\sim~$1 Gyr. 
The inferred SFHs are also used to place constraints on the properties and evolution of the ETG 
progenitors. In particular, the ETGs of our samples should have formed most stars through a phase 
of vigorous star formation (SFRs $\gtrsim350-400$ $M_{\sun}~$yr$^{-1}$) at $z\gtrsim4-5$ and  
are quiescent by $z\sim1.5-2$.
The expected number density of ETG progenitors, their SFRs and contribution
to the star formation rate density of the Universe, the location on the star formation "Main Sequence" and the required existence of massive quiescent galaxies at $z\lesssim2$, are compatible with the current observations, although the uncertainties are still large. 

}
 {
 Our results represent an attempt to demonstrate quantitatively the evolutionary
link between the most massive ETGs at $z<0.3$ and the properties of suitable progenitors at high 
redshifts. Our results also show that the full-spectrum fitting is a powerful and complementary 
approach to reconstruct the star formation histories of massive quiescent galaxies. 
 }


   \maketitle
%

\section{Introduction}

Despite the success of the $\Lambda$CDM cosmology (\citealp{Riess+1998}, \citealp{Perlmutter+1999}, \citealp{Parkinson+2012}, \citealp{Planck2015}), the complete 
understanding of how baryonic matter evolves leading to the formation of 
present-day galaxies is still unclear. The presence of cold dark matter in
the $\Lambda CDM$ framework implies a \textit{bottom-up} or \textit{hierarchical} formation of structures in which the baryonic mass growth is 
driven by the merging of dark matter halos which
progressively assemble more massive galaxies. However, recent studies showed
that merging is not the only channel to build galaxies. Depending on the 
redshift and dark halo mass, the baryonic mass can gradually grow also thanks 
to filamentary streams of cold gas (T$\sim~10^4$ K) capable to penetrate the 
shock heated medium in dark matter halos and feed the forming galaxies 
\citep{Dekel+2009}. The relative importance of these two channels of mass 
growth as a function of cosmic time is one of the key questions in order to 
explain how protogalaxies evolved into the different types of galaxies that 
we see today in the Hubble classification. 

In this context, elliptical (E) and lenticular (S0) galaxies, collectively 
called early-type galaxies (ETGs), have always been considered ideal probes 
to investigate the cosmic history of mass assembly. Indeed, holding 
the major share of stellar mass in the local Universe (\citealp{Renzini2006} 
and references therein), massive ETGs are supposed to be the endpoints of the 
hierarchical evolution, thus enclosing fundamental information about the galaxy 
formation and mass assembly cosmic history. In this regard, several 
studies attempted to infer the evolutionary properties of massive ETGs at 
$z\sim0$ ({\em archaeological} approach), or to observe directly their 
progenitors at higher redshifts ({\em look-back} approach) (see again 
\citealp{Renzini2006} for a review). The results of these studies suggest an 
empirical evolutionary trend, called {\em downsizing} (see \citet{Cowie+1996} 
for its first definition), where massive galaxies formed earlier and faster 
than lower mass systems. 

The downsizing scenario is evident in several cases of galaxy evolution 
\citep{Fontanot+2009}. In the case of ETGs at $z\sim0$, one of the first observational 
evidences can be referred to the studies of \citet{Dressler+1987}, \citet{
Faber+1992} and \citet{Worthey+1992}, who found more massive elliptical 
galaxies to be more enriched in $\alpha$-elements than less massive ones. 
These works suggested selective 
mass-losses, different initial mass functions (IMF) and/or different star 
formation timescales as possible explanations of the high level of 
[$\alpha$/Fe]. Subsequent studies found the same trend of [$\alpha$/Fe] with 
mass (\citealp[e.g][]{Carollo+1993}, \citealp{Davies+1993}, \citealp{Bender+1993}, \citealp{Thomas+2005}, \citealp{Thomas+2010}, \citealp{McDermid+2015}), 
and led to the dominant interpretation that in more massive ETGs, 
the duration of star formation was substantially shorter than in less massive 
ones, with timescales short enough (e.g. $<~$0.5 Gyr) to avoid the dilution of 
the $\alpha$ element abundance (produced by Type II supernovae) by the onset 
of Fe production by Type Ia supernovae. This is considered one of the main 
evidences of the downsizing of the star formation duration. Also the age of 
the ETG stellar populations at $z\sim0$ show evidence of downsizing, with more 
massive objects being older than less massive ones. These results have been 
derived both in clusters (\citealp[]{Thomas+2005}, \citeyear{Thomas+2010}; 
\citealp{Nelan+2005}) and in the field (\citealp{Heavens+2004}, 
\citealp{Jimenez+2007}, \citealp{Panter+2007}; see also \citealp{Renzini2006}). 
Most of these studies are based on fitting individual 
spectral features with the Lick/IDS index approach (\citealp{Burstein+1984}, 
\citealp{Worthey+1994}) which allows to mitigate the problem of the 
age-metallicity degeneracy (\citealp{Graves&Schiavon2008}; \citealp{Thomas+2005}, \citeyear{ 
Thomas+2010}; \citealp{Johansson+2012}; \citealp{Worthey+2013}). 
However, more recently, other approaches based on the full-spectrum fitting 
have been developed (e.g. STARLIGHT, \citealp{cidfernandes+2005}; VESPA, 
\citealp{tojeiro+2009}, \citeyear{tojeiro+2013}; FIREFLY, 
\citealp{Wilkinson+2015}), and applied to samples of ETGs at $z\sim0$ 
(\citealp{Jimenez+2007}, \citealp{Conroy+2014}, \citealp{McDermid+2015}). The results based on Lick 
indices are in general rather consistent with those of full spectral fitting 
within $10-30~\%$ (e.g. \citealp{Conroy+2014}) and support the downsizing
evolutionary pattern.  

In addition to the archaeological constraints, also the look-back studies
are providing complementary constraints on the evolution of the 
ETGs. First of all, massive and passive ETGs ($M\sim 10^{11}~
M_{\sun}$) have been unexpectedly discovered in substantial number 
up to $z\sim 3$ (e.g. \citealp{Cimatti+2004Nat}, \citealp{McCarthy+2004}, \citealp{Kriek+2006}; \citealp{Gobat+2012}, \citealp{Whitaker+2013}; \citealp{Onodera+2015},
\citealp{Mendel+2015}). 
For a fixed mass, these ETGs are on average more compact and therefore denser than 
present-day analogs (\citealp{Daddi+2005}, \citealp{Cimatti+2008}), especially 
at $z>1.5$ (e.g. \citealp{Cimatti+2012}). The mere existence of these massive
systems with absent star formation and old stellar ages ($\sim~1-3$ Gyr) 
implies that their star formation ceased at $z>2-3$ (consistently with the 
downsizing scenario) and that their formation mechanism was necessarily
fast and leading to a rapid assembly of compact and dense systems. 
It is relevant to recall here that at the time of their first discovery,
these massive, passive and old galaxies at $z>1.5$ were not expected at 
all in theoretical models of galaxy formation (\citealp{Cimatti+2002}, \citeyear{Cimatti+2004Nat}).

More constraints come also from the evolutionary trends emerging from 
statistical samples of ETGs at $z>0.5$ and support the downsizing picture.
For instance, the evolution of the fundamental plane indicates a decreasing 
formation redshift ($z_{F}$) for galaxies with decreasing mass both in clusters and in 
the field \citep{Treu+2005}. Moreover, if compared to the local one, the 
faint-end of the luminosity function up to $z\sim1$ is progressively 
depopulated going to higher redshift, contrary to its high-luminosity end, 
suggesting again that more massive galaxies assembled their mass earlier than 
less massive ones \citep{Cimatti+2006}. Furthermore, it has been observed 
that the galaxy number density rapidly increases from $z=0$ to $z=1$ for 
$M\lesssim 10^{11}M_{\sun}$, while it has a slower increase for more 
massive galaxies (\citealp{Pozzetti+2010}, \citealp{Moresco+2013}), starting to show a significant 
variation only at $z\gtrsim1$ for the most massive systems ($M\gtrsim 
10^{11}M_{\sun}$) (\citealp{Ilbert+2010}, see also \citealp{vanDokkum2005}, \citeyear{vanDokkum+2010}, \citealp{Muzzin+2013}).

The physical interpretation of these results on ETG evolution coming from
the archaeological and look-back approaches is not trivial in the hierarchial 
merging scenario of $\Lambda$CDM cosmology because the mass downsizing 
evolution seems anti-hierchical. However, several progresses have been made 
in the last decades. For instance, the combination of N-body simulations of
dark matter halos evolution (\citealp{Springel+2005}, \citealp{Boylan-Kolchin+2009}) with semi-analytic models for galaxy formation (\citealp{
White&Frenk1991}, \citealp{Kauffmann+1999}, \citealp{Springel+2005}, 
\citealp{Lu+2011}, \citealp{Benson2012}) have allowed a significant advance.
For example, \citet{DeLucia+2006} were able to reproduce the age-downsizing 
of elliptical galaxies by invoking AGN feedback to quench the star formation 
earlier in more massive systems with respect to less massive ones within
the standard framework of the hierarchical assembly for stellar mass. The role of AGNs in influencing 
galaxy evolution and quenching star formation is supported by several observations (\citealp{Fabian2012} and references therein, \citealp{Cimatti+2013}, \citealp{Cicone+2014}, \citealp{ForsterSchreiber+2014}); however, other models are capable to form rapidly ETGs without invoking
the AGN feedback (e.g. \citealp{Naab+2006}, \citeyear{Naab+2009}; \citealp{Khochfar&Silk2006}, \citealp{Johansson+2012_2}).
More recently, \citet{Henriques+2013}, (\citeyear{Henriques+2015}) found that 
the early build-up of low mass galaxies predicted by many models could in 
fact be prevented by assuming less massive systems to reincorporate the gas 
ejected by supernova-driven winds later than more massive objects. 
This kind of assumption, although requiring further investigation, would 
also reproduce the observed \textit{mass assembly downsizing}. However,
despite the improvements on the theoretical side, many questions remain
still open and the physics of massive galaxy formation is not fully
understood.

The empirical downsizing scenario implies that the star-forming progenitors
of massive and passive ETGs should be present at $z\gtrsim2$. Thus, the 
identification of these proto-ETGs is a crucial test to validate the
downsizing picture and to link the evolutionary properties of ETGs
inferred at $z\sim0$ with those derived directly at higher redshifts
with the look-back approach. The current knowledge on the ETG precursors
is still fragmentary and a coherent picture is lacking. However, 
star forming systems with different levels of SF activity and sometimes
with compact sizes have been identified at $z\gtrsim2-3$ as potential
star-forming progenitors of massive ETGs at $z\sim0$ (e.g. \citealp{Daddi+2004}, \citealp{Finkelstein+2013Nat},  \citealp{Williams+2015}). 
An additional piece of information comes from the host galaxies of QSOs
at $z\gtrsim6$, which have stellar masses up to $M\sim~10^{11}~
M_{\sun}$ and high metallicity (e.g. \citealp{Fan+2003}), and also from the surprising identification
of a significant number of galaxies at $z\gtrsim4-5$, which have stellar masses up to $M\sim~10^{11}~
M_{\sun}$ and look already
quiescent at these high redshifts (e.g. \citealp{Mobasher+2005}, \citealp{Wiklind+2008},
\citealp{Juarez+2009}, \citealp{Brammer+2011}, \citealp{Marsan+2015}), although their properties are based on 
photometric redshifts and SED fitting, since they are too faint
for spectroscopic identification. Should these galaxies be really
ETG analogs at $z\gtrsim4-5$, they would imply that a fraction of massive
galaxies formed rapidly at very high redshifts (say $z>5$) and with an intense star formation, if they are 
observed already quiescent at $z\sim4-5$.\\
In this work, we apply the archaeological approach to study the
properties of a sample of ETGs at $0.02<z<0.3$, restricting our analysis 
only to the most extreme systems, i.e. to the most massive ($M\gtrsim 5.6\times 10^{10}~M_{\sun}$) 
and passive objects. The aim is to infer their star formation histories and
place quantitative constraints on their precursors at high redshifts. In particular, instead of relying on the individual 
spectral features of these galaxies, we exploit the information contained in their full-spectrum, by means of the public code STARLIGHT
(\citealp{cidfernandes+2005}).\\
Throughout this work, we adopt a flat $\Lambda$CDM cosmology with $\Omega_{M}$ 
= 0.27; $H_{0}$ = 72 km s$^{-1}$Mpc$^{-1}$. Moreover, we always refer to galaxy stellar masses.
  
\section{Testing STARLIGHT with simulated ETGs spectra}
\label{sec:simulation}
\begin{figure}[!t]
\begin{center}
\includegraphics[width=1\hsize]{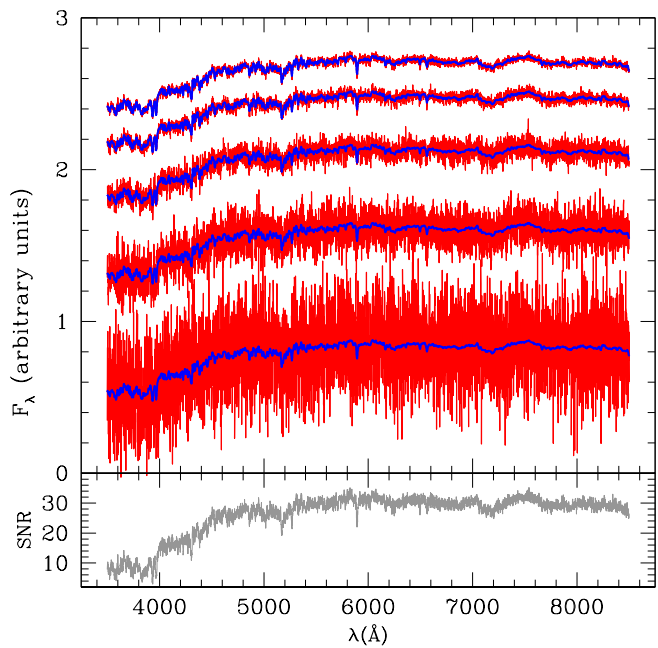}
\caption{Simulated spectra of a CSP at 13 Gyr from the onset of the SFH (exponentially delayed with $\tau$ = 0.3 Gyr), with solar metallicity and SNR 2, 5, 10, 20 and 30, increasing form bottom to top. For each SNR, the blue curve is the starting BC03 model used for the simulation, while the red curve is the the simulated spectrum. The lower panel shows the behaviour of the SNR as a function of wavelength for a simulated spectra with SNR $~=30$ in the 6500 -- 7000 \AA\  window.}
\label{fig:simul1}
\end{center}
\end{figure}

Our work is based on spectral analysis and makes use of the full-spectrum
fitting code STARLIGHT\footnote{More informations are available on the website: \textit{www.starlight.ufsc.br}} \citep{cidfernandes+2005}, which provides a fit to both galaxy spectral continuum and spectral features. \\ Basically, the best fit model derived by STARLIGHT is obtained from the combination of many SSP spectra defined in user-made libraries. The contribution of each library spectrum to the best fit model is enclosed in the so-called \textit{light-} and \textit{mass-fraction population vectors}, which contain, respectively, the light and the mass fractions with which each library model contributes to the best fit spectrum at a reference $\lambda$.
These population vectors are derived together with the best fit values for both the stellar velocity dispersion ($\sigma$) and stellar extinction ($A_{V}$).\\
Before applying STARLIGHT to real data, we test its reliability to reproduce the evolutionary properties, such as age, metallicity, star formation histories (SFHs), dust extinction and velocity dispersion of massive and passive galaxies as a function of
the signal-to-noise ratio (SNR) of the input spectra. To do this, we simulate spectra with
different SNR and known evolutionary and physical properties treating them as STARLIGHT inputs, and then we compare our assumed inputs with the code outputs.\\
We create an ensemble of simulated spectra from 3500 to 8500 \AA\ with increasing SNR, starting from BC03 (\citealp{Bruzual&Charlot2003}) theoretical models. These synthetic models are based on the STELIB stellar spectra library (\citealp{LeBorgne+2003}) and assume a  Charbrier IMF (\citealp{Chabrier2003}). Moreover, they are defined in the wavelength range $3200-9500$ \AA\ and have a resolution of $\sim~$3 \AA, which is  similar to the resolution of the SDSS spectra analyzed in the second part of this work (see Sect.  \ref{sec:sample}). We first define an error $e_\lambda$ for each theoretical spectrum by dividing the median flux of the model in the
$6500-7000$ \AA\ window by a pre-defined grid of SNR (going from 1 to 10 and from 20 to
500 with step of 10); then, we apply the $\lambda$-independent $e_\lambda$ to the
whole model spectrum, generating gaussian random numbers within
it and thus producing the simulated flux, as illustrated in Fig. \ref{fig:simul1}.
We verified that the assumption of a wavelength-independent error is reasonable, since it reproduces the behaviour of the error of individual SDSS spectra, on which we base our subsequent analysis. In particular, within the rest-frame range 3500 -- 7000 \AA\ (used to fit our data, see Sect. \ref{sec:sample}), the error of SDSS spectra is constant as a function of wavelength, apart from a discontinuity in correspondence of the overlapping of the blue and the red arm of the spectrograph (at $\sim$ 6000 \AA\ ). As a consequence, the SNR in our simulations is wavelength-dependent, as shown in Fig. \ref{fig:simul1}.\\
We assume a composite stellar population (CSP) with an exponentially delayed SFH: $\psi(t)\propto \tau^{-2}~t~exp(-t/\tau)$ and 13 different ages from the beginning of the star formation (going from 1 to 13 Gyr with step of 1 Gyr) as starting theoretical models.  We assume solar metallicity ($Z=0.02$) and a velocity dispersion $\sigma$ = 200 km s$^{-1}$. We chose a short $e$-folding time scale ($\tau=0.3$ Gyr) in order to reproduce as much as possible old ETGs spectra (\citealp{Renzini2006}). It is worth noting that our simulated spectra are also dust-free (i.e. $A_{V}=0$ mag).
To better quantify the code reliability, we also  compute 10 simulations for 
each age, thus obtaining 130 spectra for each SNR (10 simulations 
$\times$ 13 models).\\

To fit the input simulated spectra, we create a spectral library containing 176 BC03 simple stellar population (SSP) spectra. We choose  four different metallicities  for the synthetic spectra ($Z=0.004$, $Z=0.008$, $Z=0.02$, $Z=0.05$) and ages going from 0.001 to 14 Gyr. In particular, to produce a well-sampled spectral library which would help in reconstruct the SFH of the input CSPs, we define a smaller time step (0.01 Gyr) for very young ages, to take into account the very different spectral shape of stellar populations younger than 1 Gyr; a larger (0.5 Gyr) time step is instead adopted for older SSPs. We fit the input spectra within the full range $3500-8500$ \AA\  and, as required by STARLIGHT, we fix the normalization wavelength for the spectral fitting at the line-free wavelength of 5520 \AA,  thus avoiding absorption features which could influence our results. 
We perform these fits in the two cases of masking or not the input spectral regions related to emission or depending on $\alpha$-element abundances (all the masked regions are reported in Table \ref{tab:mask}), which are not modelled by the BC03 synthetic spectra. Throughout the paper, we mainly discuss the results obtained in the second case, since it is more general and easily comparable with the literature.

\begin{table}[h]
\caption{Spectral regions masked in the fit.}
\centering
\begin{tabular}{l|c} 
 \hline\\ [-1.0em] 
  & [O II]3727 [O III]5007 [O I]6302\\
Emission & H$_{\epsilon}$ H$_{\delta}$ H$_{\gamma}$ H$_{\beta}$ H$_{\alpha}$  \\ 
lines& HeI5877 \\
& [Ne III]3868 [N II]6584 [S II]6712\\
 \hline\\
$\alpha$-element& [CN1-CN2] Ca4227 G4300 \\
dependent&$Mg_{1}$ $Mg_{2}$ $Mgb$\\
lines&  Fe4383  Fe5335\\
\hline
\end{tabular}
\label{tab:mask}
\end{table}

\subsection{Definition of the evolutionary properties}
Since our tests on STARLIGHT are based on the comparison between input and output quantities, a consistent and accurate definition of them is necessary. In particular, we use mass-weighted properties, instead of light-weighted ones, since they allow to be almost independent of the choice of the normalization wavelength for the fit. Furthermore, mass-weighted quantities have the advantage to offer a more direct probe to the integrated SFH being not biased towards younger ages. Indeed, it is well-known that younger populations can dominate the luminosity of a stellar population, even when they contribute very little to the mass (\citealp{Trager+2000}, \citealp{Conroy2013}). Throughout this work, we also compare the retrieved mass-weighted and light-weighted properties (see Sect. \ref{sec:result}).

\begin{itemize}

\item \textit{Age}. Following the definition given by \citet{Gallazzi+2005} and \citet{Barber+2007}, we define the input ages as mass- and light-weighted ages, according to the following equations: $$\langle t \rangle_{mass}=\frac{\displaystyle\int_0^t\!SFR(t-t')M(t')t'\mathrm{d}t'}{\displaystyle\int_0^t\!SFR(t-t')M(t')\mathrm{d}t'} .$$\\ $$\langle t \rangle_{light}=\frac{\displaystyle\int_0^t\!SFR(t-t')f_{\lambda}(t')t'\mathrm{d}t'}{\displaystyle\int_0^t\!SFR(t-t')f_{\lambda}(t')\mathrm{d}t'} ~~,$$
 
where $M(t')$ is the stellar mass provided by an SSP of age t',  $f_{\lambda}$ is the flux at a given wavelength\footnote{Our normalization wavelength is fixed at 5520 \AA\ } of an SSP of age $t'$ and $SFR(t-t')$ is the star formation rate at the time $(t-t')$, when the SSP was formed.\\
These equations account for the fact that a composite stellar population can be treated as the sum of many SSPs of different age. Thus each SSP, with a certain age $t'$, contributes to the global mass- (light-) weighted age of the CSP according to its mass (flux) and to the SFR at the time of its formation.\\
About the output quantities, we define the retrieved output mass- and light- weighted ages by means of the mass- and light-fraction population vectors, according to the equations:

\begin{equation}
\langle t \rangle_{mass}=\sum_{j} age_{j}\cdot m_{j}~~,
\label{eq:mwa}
\end{equation}

and  

\begin{equation}
\langle t \rangle_{light}=\sum_{j} age_{j}\cdot x_{j}~~,
\label{eq:lwa}
\end{equation}

where $age_{j}$ is the age of the model component $j$, while $m_{j}$ and $x_{j}$ are its fractional contribution to the best fit model, respectively, provided by the STARLIGHT code.\\

\item{\textit{Age of formation}.} Starting from the output mass-weighted ages, we define the "age of formation" $age_{f}$ (i.e. look-back time) of the analyzed galaxies as:

\begin{equation}
\label{eq:agef}
age_{f}=age_{model}+\Delta t~~,
\end{equation}

where $age_{model}$ is the original age of each model in our spectral library and $\Delta t=[age_{U}(z=0)-age_{U}(z)]$ is the difference between the age of the Universe today and the age of the Universe at the redshift of the galaxy. In
other words, $age_f$ allows to consistently scale all the derived ages to
$z=0$. \\

\item \textit{Metallicity}. As said above, we consider solar metallicity
 ($Z=0.02$) for the input simulated spectra; we define the output metallicities only as mass-weighted metallicities, employing the following equation:

\begin{equation}
\langle Z \rangle_{mass}=\sum_{j} Z_{j}\cdot m_{j}~~,
\label{eq:mwz}
\end{equation}

where $Z_{j}$ is the metallicity of the model component $j$.\\

\item \textit{Star formation histories.} In general,  considering that a CSP can be viewed as the sum of many SSPs (see eq. [1] in \citealp{Bruzual&Charlot2003}), the SFH of a stellar population can be understood as the fraction of stellar mass produced as a function of time in the form of SSPs. For this reason, the mass fractions $m_{j}$ provided by STARLIGHT plotted as function of the library SSPs ages can be considered as a direct proxy for the output SFH.
Starting from this, we use the following equation to define the input SFH: $$SFH_{in}=\frac{SFR(t-t')M(t')\mathrm{d}t'}{\displaystyle\int_0^t\!SFR(t-t')M(t')\mathrm{d}t'}~~,$$ where $\mathrm{d}t'$ was introduced to make $SFH_{in}$ dimensionless, just like the output $m_{j}$.\\

\item \textit{Extinction and velocity dispersion}. Extinction and velocity dispersion are direct STARLIGHT output, so we compare them directly with the chosen input ones (i.e. $A_{V}=0$ mag and $\sigma$ = 200 km s$^{-1}$).
\end{itemize}

\subsection{Results from the simulations}
\label{sect:sim}

\subsubsection{Age and metallicity}
The two top panels of Fig. \ref{fig:fusion} illustrate the results concerning ages and metallicities.
In this figure we show the median differences, averaged on all the realized simulations at each SNR, between the output and the input quantites as a function of the mean SNR of the simulated spectra, together with their median absolute deviation (MAD)\footnote{Given an ensemble of data $X_{i}$, the median absolute deviation is defined as: MAD$~=1.48 \times median(\lvert X_{i}-median(X_{i})\rvert$, see \citet{Hoaglin1983}.}, calculated at each SNR on the available simulations. We found that the differences and the relative dispersions effectively decrease with increasing SNR, as expected. In the case of the ages, only a small systematic shift of about +0.3 Gyr is present at SNR $\gtrsim~$10. To better quantify our results, we defined the percentage accuracy as the ratio between the dispersion of a given quantity on all the simulations at a given SNR and its true value, calculating the minimum SNR at which the input properties are retrieved with a percentage accuracy higher than $\sim~$10$~\%$. \\ We find minimum SNR values of $\sim~$18 for ages and $\sim~$9 for metallicities. Table \ref{tab:shiftdisp} lists the median shifts and the median dispersions (together with their percentage values) 
which we find in the  case of SNR$~=100$, 
which is close to the typical SNR of the median stacked spectra which we are going to analyze in Sect. \ref{sec:sample} (results for SNR$~=~$20 are also shown to facilitate the comparison with literature).\\ 
We find that, masking the spectral regions associated with emission or depending on $\alpha$-element abundances, the SNR needed to reach a percentage accuracy larger than the 10$\%$ increases to $\sim~$40 for mass-weighted ages, remaining unchanged for the other quantities. We ascribe this effect to the fact that we are exluding from the fit some spectral regions which are well-known age indicators (e.g. the $H\beta$ line, see \citealp{Burstein+1984}, \citealp{Worthey+1994}).

\subsubsection{Dust extinction and velocity dispersion}
\begin{figure}[!t]
\begin{center}
\includegraphics[width=1\hsize]{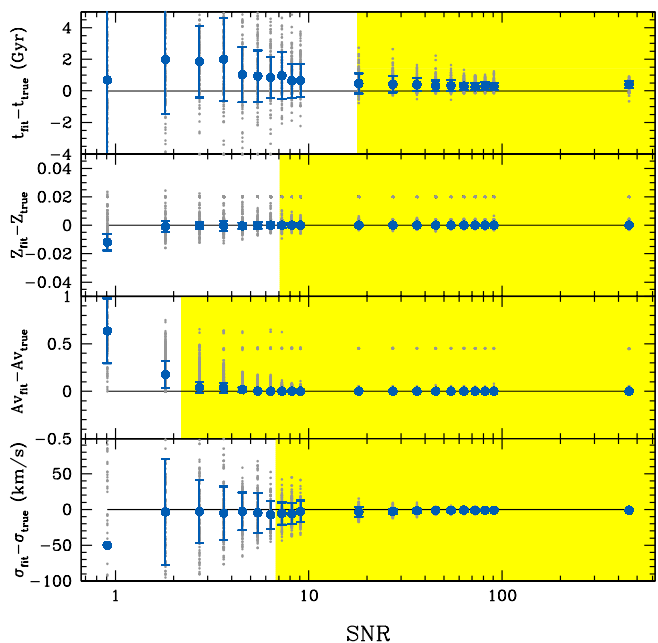}
\caption{Retrieved ages, metallicities, dust extinctions and velocity dispersions as a function of the mean SNR of the input spectra. Blue filled circles are the differences, averaged on all the computed simulations (i.e. 10$~\times~$13 models), between the ouptut and the input mass-weighted ages, mass-weighted metallicities, dust extinctions and velocity dispersions (from top to bottom); blue vertical bars are the median absolute deviations (MAD) on all the computed simulations; grey points are the differences derived from each individual simulation;
yellow shaded regions in each panel indicate the minimum SNR from which the accuracy of the input retrieving is better than 10$~\%$.}
\label{fig:fusion}
\end{center}
\end{figure}

Dust extinction and velocity dispersion are retrieved with a percentage accuracy higher than the 10$~\%$ for SNR$~\gtrsim~$7 and $\gtrsim~$3, respectively, as illustraded in the two lowest panels of Fig. \ref{fig:fusion} and reported in Table \ref{tab:shiftdisp}. 
The results are in agreement with the ones obtained by \citet{Choi+2014}, who used another full-spectrum fitting code (developed by \citealp{Conroy&VanDokkum2012a}) and by \citet{Magris+2015}, who adopted the STARLIGHT code itself. Indeed, these authors found that age, metallicity, velocity dispersion and dust extinction are recovered without significant systematic offsets starting from SNR$~\gtrsim~$10. In particular, for a SNR of 20, their median percentage shifts are  $\lesssim10~\%$ on age  and metallicity (\citealp{Choi+2014}) and $\sim~$8$~\%$ on $\sigma$ (\citealp{Magris+2015}), while their dispersions are $\lesssim10~\%$ on age and metallicity (\citealp{Choi+2014}) and $\sim~$15$~\%$ on $\sigma$ (\citealp{Magris+2015}). All these values are in agreement with ours at the same SNR (see Table \ref{tab:shiftdisp}).\\ 
It is worth noting that, if we mask the spectral regions listed in Table \ref{tab:mask}, similar SNR of $\sim3$ and $\sim8$ are needed to recover dust extinction and velocity dispersion with a percentage accuracy better than $10~\%$, respectively. 

\subsubsection{Star formation histories including a single CSP}
\begin{figure}[!t]
\resizebox{\hsize}{!}{\includegraphics{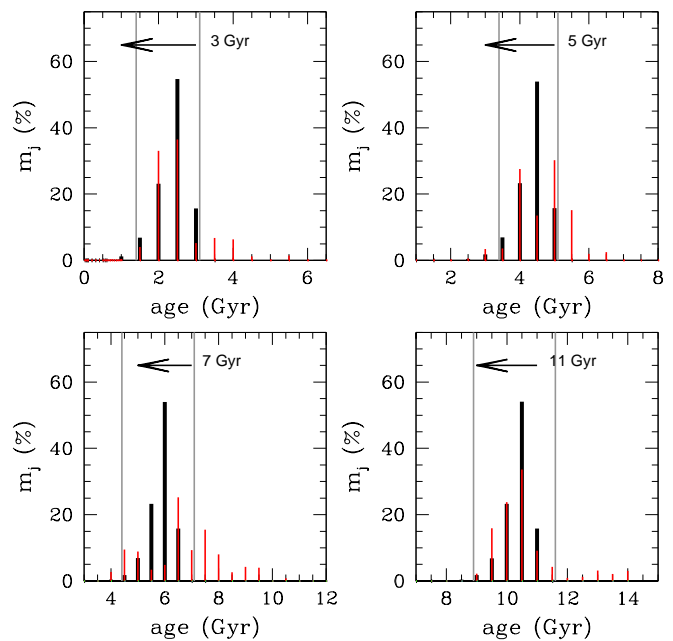}}
\caption{Recovery of star formation histories. The four panels show a CSP at 3, 5, 7 and 11 Gyr from the onset of the SFH. Black vertical lines represent the exponentially delayed SFH ($\tau=0.3~$Gyr) assumed for the input spectra (rebinned according to the age step of our library models), while vertical lines represent the output SFH; horizontal arrows indicate the time from the beginning of the SF (i.e. 3, 5, 7 and 11 Gyr, respectively), while grey vertical lines mark the asymmetric age ranges around the SFH peak defined in the text for CSPs younger or older than 5 Gyr (i.e. [$-~1$; $+~0.5$] and [$-~1.5$; $+~1$], respectively).}
\label{SFHretrieving1}
\end{figure}

\begin{figure}[!t]
\resizebox{\hsize}{!}{\includegraphics{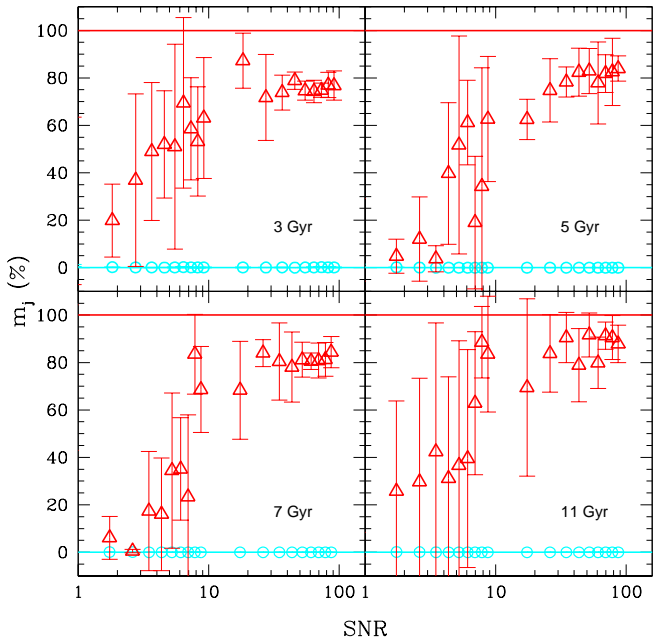}}
\caption{Recovery of star formation histories. The four panels show a CSP at 3, 5, 7, and 11 Gyr from the onset of the SFH. In the two top panels, red triangles indicate the median mass fractions retrieved within [$-~1$; $+~0.5$] Gyr from the SFH peak; in the two bottom panels, they are the median mass fractions within [$-~1.5$; $+~1$] Gyr (as described in the text). In all four panels, cyan circles are the median mass fractions relative to stellar populations younger than 0.5 Gyr, while red and cyan horizontal lines mark the expected mass fraction within the defined age ranges and below 0.5 Gyr, respectively.}
\label{SFHretrieving2}
\end{figure}

To quantify the accuracy with which the SFHs are retrieved, we calculate the percentage of mass gathered around the peak of the SFH, comparing the results with the theoretical expectations, as illustrated in Fig. \ref{SFHretrieving1}. Here, the case of a SNR$~=~$50 CSP input spectrum at four different ages from the onset of the SFH is shown.  In particular, we find that the theoretical exponentially delayed SFH assumed for the input spectra (rebinned at the age step of our SSP library models) is narrow, encompassing $\sim~$100$~\%$ of mass within an age interval going from $-~1$ Gyr to $+~0.5$ Gyr around its maximum. In comparison, for SNR $\gtrsim~$10, the retrieved SFHs appear slightly broader with respect to the theoretical one. Indeed, $\sim~$80$~\%$ of mass is gathered within [$-~1$; $+~0.5$] Gyr for CSPs younger than 5 Gyr, and an even larger age interval of [$-~1.5$; $+~1$] Gyr is necessary to get this same mass percentage in the case of older populations. This dependence is illustrated in a more general way in Fig. \ref{SFHretrieving2}, which shows the median mass fractions gathered around the SFH peak as a function of the mean SNR of the simulated spectra, for four CSPs. 
It is possible to note that CSPs older than 5 Gyr require larger age intervals around the SFH maximum to gather the same mass percentage as younger stellar populations. However, since a significant percentage of mass ($70-80~\%$) is always reached within $\sim~$1 Gyr from the peak of the SFH at any age, we conclude that the full-spectrum fitting succeeds in reproducing the input SFH for SNR$~\gtrsim~$10. Finally, we find that STARLIGHT does not have any bias toward young ages, since, at any SNR, no mass is gathered at ages younger than 0.5 Gyr, except in the case of the youngest CSP (these results are valid also when the mask is applied to the input spectra).
To summarize,  from our tests we can conclude that the full-spectrum fitting with STARLIGHT is reliable if SNR $~\gtrsim~10$ -- 20 are considered. These SNRs are lower or equal than the ones of typical SDSS individual spectra (i.e. $\sim~$20) and well below the typical SNR of stacked spectra, which are mainly used for high redshift studies (see Sect. \ref{subsect:stacked}). \\

\begin{table*}[h]
\caption{Median shift and dispersion from the simulation of ETGs spectra, for SNR$~=20$ and 100 and BC03 spectral synthesis models.}
\renewcommand{\arraystretch}{1.4}
\centering
\begin{tabular}{ccccccccc} 
 \hline
 SNR & \multicolumn{2}{c}{$\langle t\rangle_{mass}$ (Gyr)}& \multicolumn{2}{c}{Z}& \multicolumn{2}{c}{$A_{V}$ (mag)}& \multicolumn{2}{c}{$\sigma$ (km s$^{-1}$)}\\
&shift & disp & shift & disp & shift&disp&shift&disp \\ 
\hline
20&0.5 (10$~\%$)&0.6 ($10~\%$)&$\lesssim10^{-3}$ ($\lesssim 1~\%$) & $\lesssim10^{-3}$ ($3~\%$)&$\lesssim10^{-3}$&$\lesssim 10^{-3}$&$-3.3$ ($4~\%$)&6.6 ($7~\%$)\\
100&0.3 (3$~\%$)&0.3 ($3~\%$)&$\lesssim10^{-3}$ ($\lesssim 1~\%$) & $\lesssim10^{-3}$ ($\lesssim1~\%$)&0.0001&$\lesssim 10^{-3}$&$-1.02$ ($2~\%$)&1.3 ($1~\%$)\\
\hline
\end{tabular}
\label{tab:shiftdisp}
\end{table*}

\subsubsection{More complex star formation histories}
\begin{figure}[!t]
\resizebox{\hsize}{!}{\includegraphics{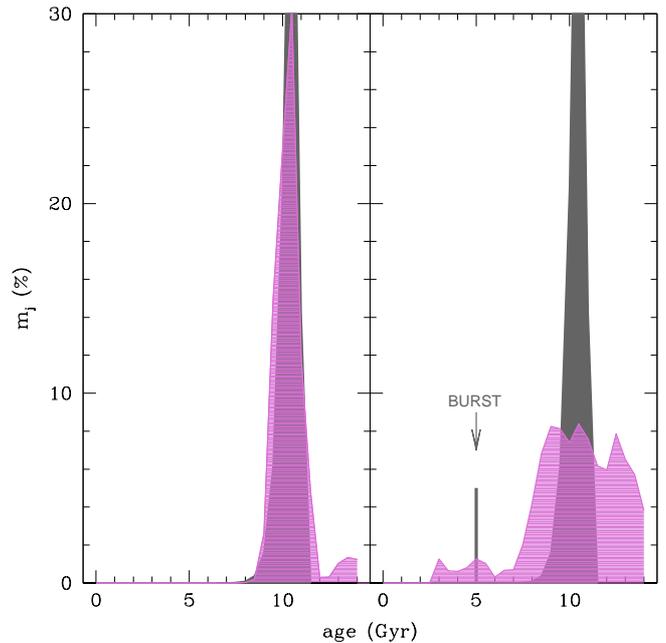}}
\caption{Retrieving of more complex SFH. Left panel: output SFH (pink curve)  obtained when an input SFH  (grey curve) including only a single CSP is considered; right panel: output SFH (pink curve) obtained when an input SFH (grey curve and vertical line) including a single CSP plus a later burst  (which happens 6 Gyr after the main SF episode and contributes to the 5 $\%$ in mass) is considered.
}
\label{SFHcfr}
\end{figure}

\begin{figure}[!t]
\resizebox{\hsize}{!}{\includegraphics{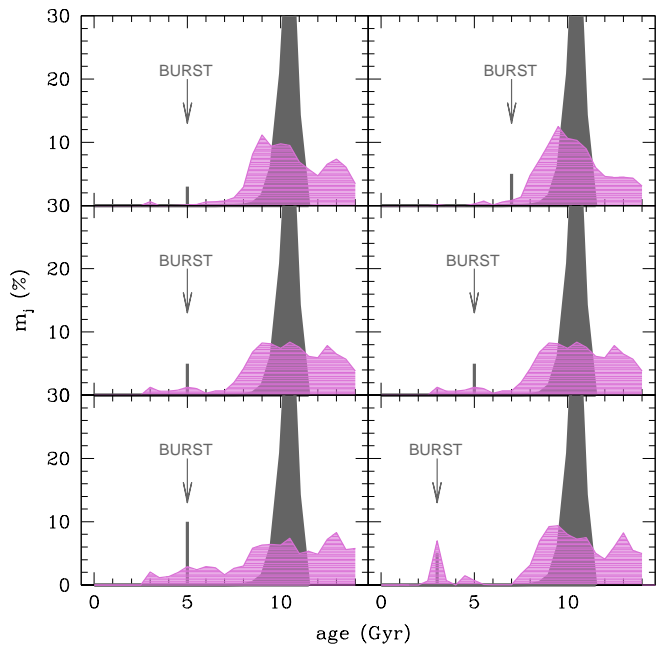}}
\caption{Retrieving of more complex SFHs as a function of the age and the mass contribution of the later burst. Left panels show the output SFH (pink curves) obtained when the input SFH (grey curves and vertical lines) includes a single CSP plus a later burst happening 6 Gyr after the main SF episode and contributing in mass by 3, 5 and 10 \% (from top to bottom). Right panels show the output SFH (pink curves) obtained when the input SFH (grey curves and vertical lines) includes a single CSP plus a later burst of SF contributing in mass by 5 \%, and happening  4, 6, 8 Gyr after the main SF event (from top to bottom).}
\label{SFHcfr2}
\end{figure}

Since the star-formation histories of galaxies (ETGs included, e.g. \citealp{DeLucia+2006}, \citealp{Maraston+2009}) can be stochastic and include multiple bursts, we also verify the full-spectrum fitting capabilities to retrieve more complex SFHs. In particular, we take an 11 Gyr old composite stellar population with an exponentially
delayed SF ($\tau=0.3$ Gyr) as the main SF episode (this age is compatible with the age of the Universe at $z\sim0.15$, which is the median redshift of our sample, see Sect. \ref{sec:sample}). We then define more complex SFHs by combining this single CSP with a burst of SF at different ages (5, 6, 7 Gyr) and with different mass contributions (3, 5, 10 $\%$). In all cases, we consider a solar metallicity for the main SF episode and, according to the results of \citet{Maraston+2009}, a subsolar metallicity ($Z=0.004$) for the later one. We do not mask any spectral feature of the input spectra, we assume $A_{V}~=~0.1$ mag for the two components and apply a velocity dispersion of 200 km s$^{-1}$. We show the results for a SNR of $80$, which matches the typical SNR of the SDSS median stacked spectra analyzed in the following (see Sect. \ref{sec:sample}).
Useful information can be derived from the comparison between the output SFH obtained from these input simulated spectra and the one provided when the single CSP alone is taken as input SFH.
Fig. \ref{SFHcfr} shows that the single CSP alone is well recovered by the full spectrum fitting. In particular, $\sim80~\%$ of the stellar mass is retrieved within $\sim1$ Gyr from the SFH peak. When a burst is added to this major episode of SF, the full-spectrum fitting is able to recognize the presence of a more complex SFH, as indicated by the tail appearing at smaller ages, and the total mass percentage of the later burst is retrieved within 1 Gyr from the expected age. However, we note that the main episode of SF is spread on a time interval longer than expected, and $50~\%$ of the stellar mass is retrieved around $\sim1$ Gyr from the SFR peak.
We also find that, in this case, the mean properties of the global stellar population are well retrieved, with a percentage accuracy larger than $10~\%$ starting from SNR$~\sim15$ for age, $ \sim7$ for metallicity, $\sim20$ for $A_{V}$,  $\sim8$ for $\sigma$ and that the metallicities of the two SF episodes are separately recovered. These SNRs are well below the typical ones of the stacked spectra analyzed in the following Sections.\\
In Fig. \ref{SFHcfr2}, we analyze the recovery of more complex SFHs as a function of the age and the mass contribution of the later SF burst. In particular, we find that the SFH tail, indicative of the presence of more recent SF episodes, arises for bursts happening $\gtrsim$ 6 Gyr after the main SF event and contributing in mass by $\gtrsim5~\%$. Therefore, under these conditions, the full-spectrum fitting is able to recognize the SFH complexity. 
Furthermore, if mergers have played a role in the galaxy star formation history, these tests suggest that our method would be sensible to variations in SFR produced by mergers among galaxies of different age,  but it is intrinsically not able to discern whether or not mergers among galaxies of similar age (coeval mergers), which increase stellar masses and sizes leaving the shape of the SFH unchanged, occurred during the galaxy evolution to build galaxy mass.

\subsection{Testing different stellar population synthesis models}
\label{appendix1}
\begin{figure}[t!]
\begin{center}
\includegraphics[width=1\hsize]{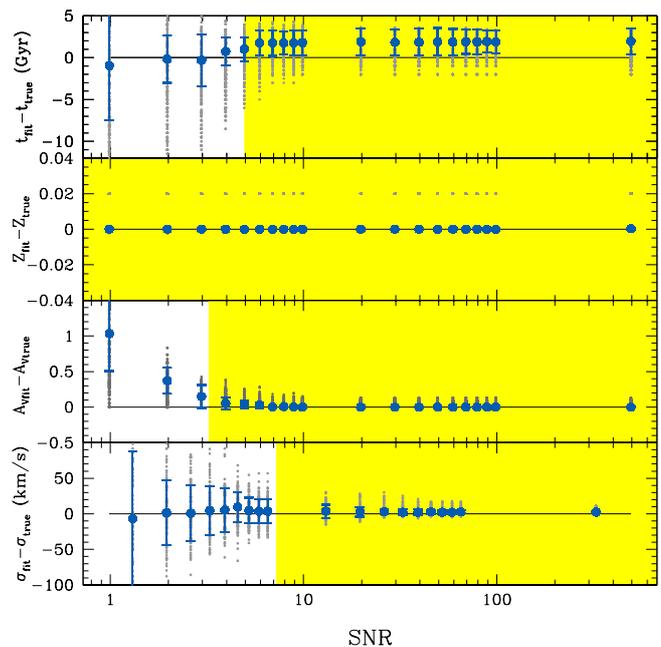}
\caption{The same of Fig. \ref{fig:fusion}, but using the MS11 spectral library to retrieve the evolutionary properties.}
\label{fig:mastro}
\end{center}
\end{figure}

Having adopted BC03 models to set up both the input simulated spectra and the spectral library, in this Section we verify if this choice can affect the results. This is a simple consistency check to verify the full-spectrum fitting capabilities to retrieve the galaxy evolutionary properties starting from the same input spectra. With this aim, using the procedure described in Sect. \ref{sec:simulation} we simulate, starting from BC03 spectra, 13 simple stellar population (SSP) spectra with solar metallicities and ages going from 1 to 13 Gyr with step of 1 Gyr, within the same SNR range defined in Sect. \ref{sec:simulation}, and we fit them with a spectral library of \citet{Maraston&Stromback2011} (MS11) SSPs models, which are defined in the wavelength range $3200-9300$ \AA, are based on the STELIB stellar spectra library \citep{LeBorgne+2003}, have a resolution of 3 \AA, and assume a Kroupa IMF \citep{Kroupa2001}. We show in Fig. \ref{fig:mastro} the recovery of age, metallicity, velocity dispersion and dust extinction, in analogy with Fig. \ref{fig:fusion}. Also in this case, the percentage accuracy of 10$~\%$ is reached at similar SNR,  which are $\sim~$5 for age, $\sim~$7 for velocity dispersion and $\sim~$3 for dust extinction, while metallicity is retrieved with basically no bias at every SNR. 
However,  as it is possible to note, a difference between the retrieved mass weighted ages and the true ones is present also at high SNR.  This overestimation can be explained by the fact that, due to a different treatment of the TP-AGB phase and of the convective overshooting in the stellar interiors, MS11 models are on average bluer than BC03 ones (see \citealp{Maraston+2006} for further details). This implies that, for the same metallicity, older MS11 models are chosen to best-fit BC03 input spectra, to compensate the bluer colors.\\


To avoid biases against MS11 models, we also  fit an ensamble of 13 MS11 simulated spectra with a spectral library made up of MS11 models themselves (this test is performed using MILES - see \citealp{SanchezBlazquez+2006} - stellar spectra both for the input and the library synthetic models). For computational feasibility, we create in this case a less extended grid of SNR for the input simulated spectra and, to be consistent with the results shown for BC03 models, we do not adopt the mask for the input spectra. In this case, we find that the evolutionary properties are retrieved with a percentage accuracy larger than the 10 $~\%$ starting from a SNR of $\sim25$ for mass-weighted ages, $\sim15$ for metallicities, $\sim30$ for dust extinction and $\sim6$ for velocity dispersion. These values are in general higher than the ones found using BC03 models but, however, it is important to note that they are below the ones of the median stacked spectra analyzed in the following Sections (see Sect. \ref{sec:sample}).

\section{Sample selection}
\label{sec:sample}

We analyze a sample of 24488 very massive and passive early-type galaxies with $0.02\lesssim z \lesssim0.3$. The restriction to the most passive ETGs is indeed very useful to investigate galaxy evolution, since it allows to address the galaxy mass assembly history in the very early epochs.\\ The adopted sample was already used by \citet{Moresco+2011} (M11) (see this work for more details), who matched SDSS-DR6 galaxies with the Two Micron All Sky Survey (2MASS) (with photometry in the J, H and K bands) to achieve a larger wavelength coverage, thus obtaining better estimates of stellar masses from SED-fitting and photometric data. According to the criteria explained in M11, the analyzed galaxies have no strong emission lines (rest-frame equivalent width EW(H$\alpha$)$>-5$ \AA\  and  EW ([OII]$\lambda3727)>-5$ \AA\ ) and spectral energy distribution matching the reddest passive ETG templates, according to the \citet{Ilbert+2006} criteria. Finally, the sample was cross-matched with the SDSS DR4 subsample obtained by \citet{Gallazzi+2005}, for which stellar metallicity estimates have been performed by means of a set of Lick absorption indices (see Sect. \ref{subsect:met}).
Moreover, to deal only with the most massive galaxies, only objects with $log(M/M_{\sun})>10.75$  were included in the sample (above this threshold, the galaxy number density is found to be almost constant up to $z\sim1$, see \citealp{Pozzetti+2010}) which, in addition, was splitted in four narrow mass bins
$(\Delta log(M/M_{\sun}= 0.25)$: $10.75 < log(M/M_{\sun} ) < 11$, $11 < log(M/M_{\sun}) < 11.25$, $11.25 < log(M/M_{\sun} ) < 11.5$ and $log(M/M_{\sun} ) > 11.5$ (Table \ref{tab:percentiles} reports the number of objects contained in each mass bin), in turn divided into various redshift bins of width $\Delta z\sim~$0.02. It is worth noting that, on average, the difference of the median mass along the redshift range is negligible within the four mass slices in which our sample is divided, showing no significant evolution as a function of redshift (\citealp{Moresco+2011}, Concas et al. 2016, in prep). \\We also assess the morphology of the galaxies in our sample using the S\'ersic index $\rho_{n}$ \citep{Sersic1968}, which correlates with the morphology because objects with $n<2.5$ are disky \citep{Andredakis+1995}, while those with $n>2.5$ are bulge-dominated or spheroids \citep{Ravindranath+2004}. In particular, the values of $\rho_{n}$ are taken from the NYU Value-Added Galaxy catalogue \citep{Blanton+2005b}, which contains photometric and spectroscopic information on a sample of $\sim2.5\times10^7$ galaxies extracted from the SDSS DR7. From this check, we derive that $\sim96~\%$ of the analyzed galaxies are bulge-dominated systems.

\subsection{Median stacked spectra}
\label{subsect:stacked}
\begin{figure*}
\centering\includegraphics[width=1\textwidth]{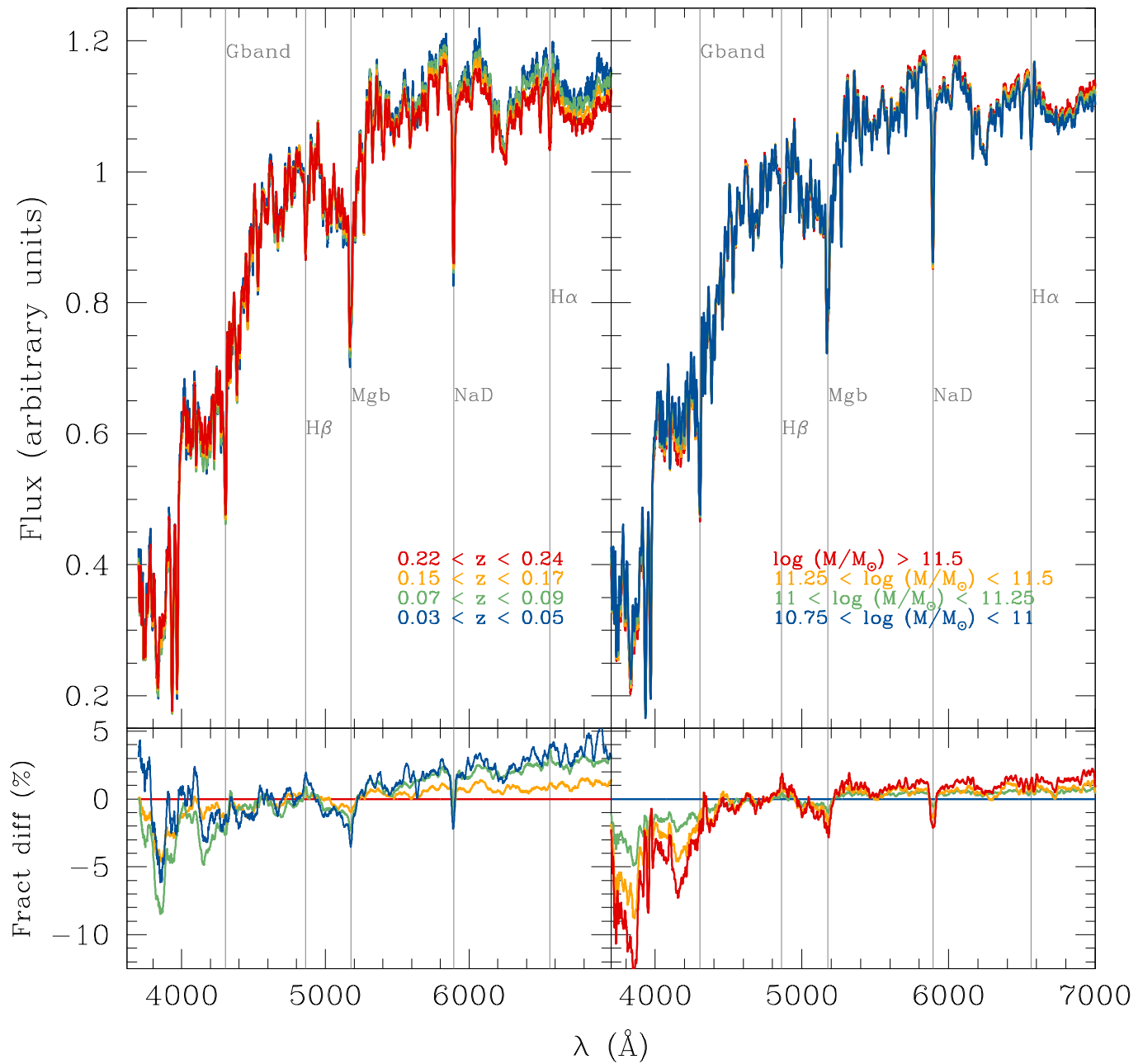}
\caption{SDSS median stacked spectra for the sample of massive and passive ETGs. Left and right upper panels show, respectively, median stacked spectra with a fixed redshift ($0.15\lesssim z \lesssim0.19$) and different masses (with mass increasing from blue to red)  and with a fixed mass ($11.25<log(M/M_{\sun})<11.5$) and four different redshifts (0.04, 0.08, 0.16, 0.23) (with redshift increasing from blue to red). Lower panels illustrate the fractional differences (defined as $(f_{i}-f_{REF}/f_{REF})\times100$, where $f_{i}$ is the flux of the i-th spectrum and $f_{REF}$ is the reference one) among the stacked spectra as a function of wavelength. In particular, to show the redshift and the mass dependence we used as reference, respectively, the median stacked spectrum obtained for $11.25<log(M/M_{\sun})<11.5$ and $0.07<z<0.09$, and the one corresponding to $10.75<log(M/M_{\sun})<11$ and $0.15<z<0.17$. Vertical grey lines mark some of the best-known absorption lines.
}
\label{fig:stackedsp}
\end{figure*}

In order to increase the SNR of the input spectra, we work on the median stacked spectra derived for each mass and redshift bin. A stacked spectrum is produced by shifting each individual rest-frame spectrum, normalized to a chosen wavelength, to the rest frame and then by constructing, via interpolation, a grid of common wavelengths at which the median (or mean) flux is computed. In our sample, the median stacked spectra were obtained for each mass and redshift bin (see Sect. \ref{sec:sample}).
We define the error on the median stacked flux as MAD$/\sqrt{N}$, were MAD is the median absolute deviation and N is the number of objects at each wavelength. Before stacking them together, we normalized each individual spectrum at rest-frame 5000 \AA, where no strong absorption features are present  which could influence the final result. It is worth noting that the error on the final stacked spectrum was corrected in order to obtain more reliable values for the $\chi^{2}$ distribution provided by the spectral fit. In particular, we readjusted the original error $e_{\lambda}$ of the spectra making it comparable with the data dispersion of $(O_{\lambda}- M_{\lambda})$ (where $O_{\lambda}$ are the observed data, $M_{\lambda}$ is the best fit model obtained from the first-round fit). After the error re-definition, the signal-to-noise ratios of the median stacked spectra are $\sim~$80 (the error $e_{\lambda}$ is, on average, increased by a factor of $\sim7$).\\
The main characteristics of the analyzed median stacked spectra are illustrated in Fig. \ref{fig:stackedsp}, where also the fractional differences among the various spectra are shown to better visualize their differences as a function of wavelength. As it is possible to note, median stacked spectra are completely emission lines-free. Moreover, two clear observational trends are present, with spectra getting redder both with cosmic time and increasing mass. In the following, our aim is to investigate these observed trends by means of the full-spectrum fitting technique. In Sect. \ref{stacked_individual}, we also verify that the use of stacked spectra does not introduce biases in the results, concluding that they can be employed at high redshift to increase the SNR of the observed spectra.

\section{Evolutionary properties from the full-spectrum fitting of median stacked spectra}
\label{sec:result}
\begin{figure*}
\includegraphics[width=1\hsize]{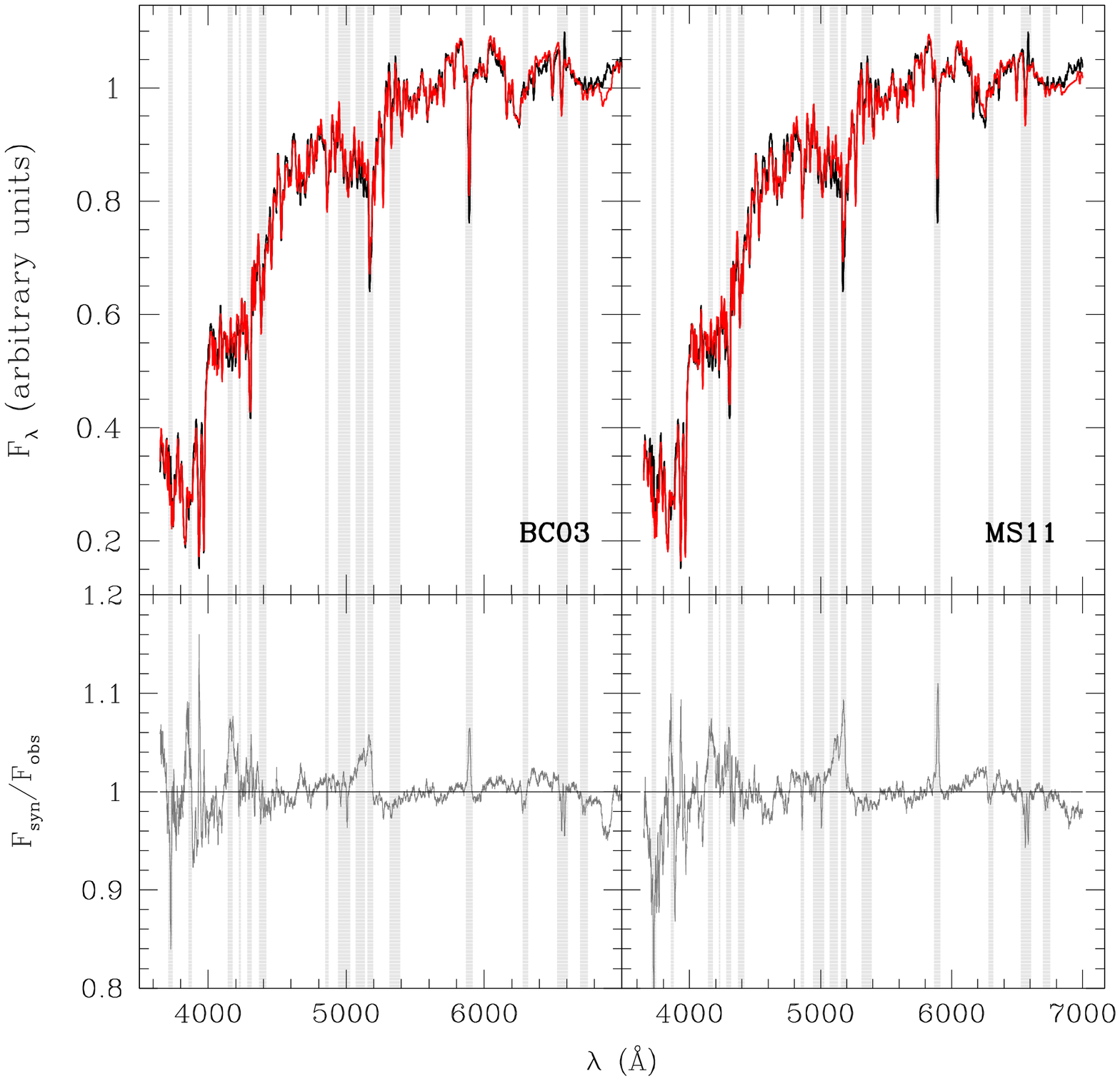}
\caption{Typical output of the full-spectrum fitting procedure using BC03 (left) and MS11 (right) spectral synthesis models. In particular, we show the case of the median stacked spectrum derived for $11<log(M/M_{\sun})<11.25$ and $z\sim0.05$. In the top panel the black curve is the observed spectrum, the red curve is the best fit model and grey shaded regions are the masked spectral regions (see Table \ref{tab:mask}). In the bottom panel the ratio of the best fit model spectrum to the observed flux is shown.}
\label{fig:riass_stacked}
\end{figure*}

To infer the evolutionary and physical properties of our sample, we fit the median stacked spectra for each mass and redshift bin, mainly using BC03 spectra, but also exploiting a spectral library of MS11 STELIB models. In all cases, we fix the wavelength range for the spectral fitting to $3500-7000$ \AA\ for all the mass and redshift bins, in order to avoid redshift-dependent results.Both the BC03 and the MS11 spectral libraries contain spectra with ages starting from 0.01 Gyr and not exceeding the age of the Universe at the redshift of the stacked spectrum itself.\\We use the metallicities $Z=0.004, 0.008, 0.02, 0.05$ for the BC03 library and $Z=0.01, 0.02, 0.04$ for the MS11 one and we assume a Calzetti attenuation curve to account for the presence of dust \citep{Calzetti2001}. Moreover, contrary to what did in Sect. \ref{sec:simulation}, in both cases we mask the spectral regions of the input spectra which are associated with emission lines \citep{cidfernandes+2005} and $\alpha$-element dependent lines \citep{Thomas+2003}, since they are not modelled by BC03 and MS11 synthetic spectra (see Table \ref{tab:mask} and grey shaded regions in Fig. \ref{fig:riass_stacked}).\\  As it is possible to note from Fig. \ref{fig:riass_stacked}, the best fit model produced by the spectral fitting matches correctly the input spectrum, both in the case of BC03 and MS11 models. 
The dispersion on the mean ratio between the synthetic and observed flux on all the not-masked pixels is $\lesssim2\%$. \\
From the comparison between the observed and the synthetic spectra, signs of a residual emission in correspondence of the masked spectral regions can be noticed.  We discuss this residual emission in a parallel paper, which is dedicated to the study of the $H\beta$ absorption line (Concas et al. 2016, in prep.).


\subsection{Error estimates}
\label{subsect:errest}

Throughout the data analysis, the uncertainties on the investigated evolutionary and physical properties are derived in three different ways, defined as follows:

\begin{itemize}

\item \textbf{$rms_{sim}$}: this is the dispersion derived in Sect. \ref{sect:sim} at the typical  SNR of the analyzed median stacked spectra (i.e. SNR$\sim~$100);\\

\item \textbf{$err_{100}$}: this error is derived by comparing the results obtained from computing 100 realizations of each median stacked spectrum within its error (defined in Sect. \ref{subsect:stacked});\\

\item \textbf{$rms_{set}$}: this dispersion is calculated by comparing the results obtained by changing some setting parameters of the spectral fit (e.g. the number of the masked spectral regions, the upper limit for dust extinction, the extension of the library in age or metallicity).

\end{itemize}

\subsection{Ages}
\label{subsect:ages}

\begin{figure}[!t]
\begin{center}
\includegraphics[width=1\hsize]{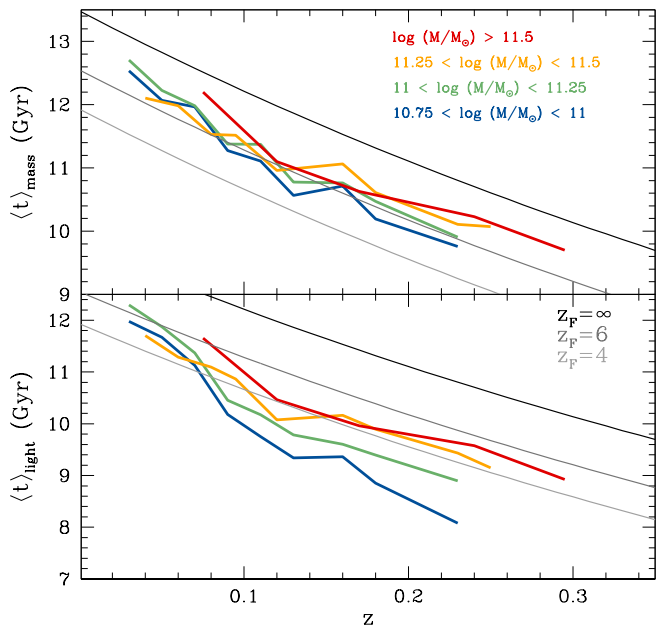}
\caption{Mass-weighted (top) and light-weighted (bottom) age-redshift relations (for BC03 models). Stellar mass increases from blue to red. The black line is the age of the Universe, while grey lines are the age of galaxies assuming different formation redshifts.}
\label{fig:zage}
\end{center}
\end{figure}

\begin{figure}[!t]
\begin{center}
\includegraphics[width=1\hsize]{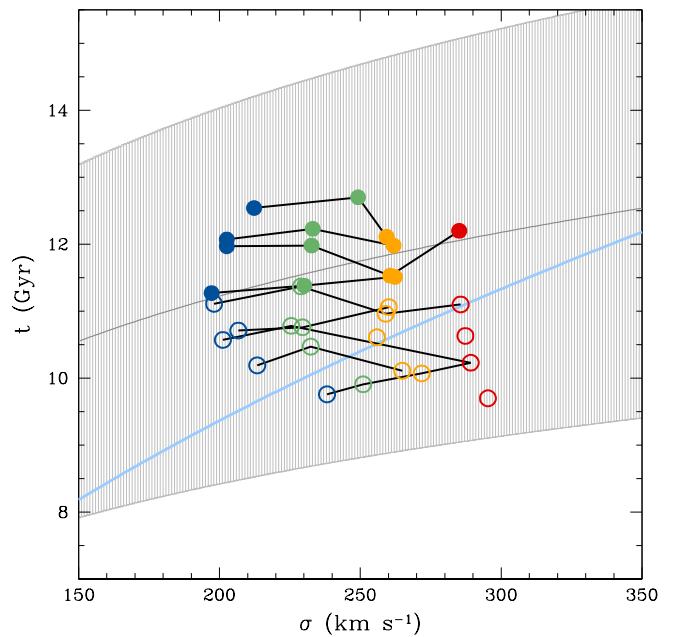}
\caption{Mass-weighted ages as a function of the velocity dispersion $\sigma$ for BC03 models (symbols are color-coded as in Fig. \ref{fig:zage}). 
Filled circles are the mass-weighted ages corresponding to $z<0.1$, matching the redshifts analyzed by \citet{McDermid+2015}, with black curves linking the mass-weighted ages related to different mass bins but similar redshifts. The grey curve is the age -- $\sigma$ relation inferred by \citet{McDermid+2015} with its dispersion (grey shaded region), while the light-blue curve is the \citet{Thomas+2010} relation (its dispersion, not shown in the figure, is of the order of $\sim~60~\%$).}
\label{fig:mcdermid}
\end{center}
\end{figure}

Using eq. (\ref{eq:mwa}), we derive the mass-weighted ages of our galaxies, illustrated in the upper panel of Fig. \ref{fig:zage}.
Considering the observed properties of the median stacked spectra with mass and redshift illustrated in Fig. \ref{fig:stackedsp}, we find an evolutionary trend, with mass-weighted ages increasing sistematically with mass (for a fixed redshift), as well as with cosmic time (for a given mass bin). In particular, ages vary from $\sim~$10 to $\sim~$13 Gyr, increasing by $\sim~$0.4 Gyr from the lowest to the highest masses, in agreement with the observational trends shown in Fig. \ref{fig:stackedsp}. The lower panel of Fig. \ref{fig:zage} illustrates, instead, the retrieved light-weighted ages, which instead increase by $\sim~$1 Gyr from the lowest to the highest mass. Light-weighted ages are younger than mass-weighted ones, with a difference of $\sim~$0.6 Gyr in the lowest redshift bins and of $\sim~$1 Gyr in the highest ones, on average (Table \ref{tab:unc} lists the three uncertainties estimates $rms_{sim}$, $err_{100}$ and $rms_{set}$ derived for $<t>_{mass}$).\\ It is interesting to note that, even when the cosmological constraint is relaxed and the maximum allowed age for the spectral library models is pushed  to 14 Gyr at all redshifts, the trend of the mass-weighted age with both mass and cosmic time still holds, and only in few cases the mass-weighted ages exceed the age of the Universe.
Fig. \ref{fig:mcdermid} illustrates our mass-weighted ages as a function of the velocity dispersion (see Sect. \ref{subsect:veldisp}). At $z<0.1$, we find that they are compatible, within the dispersion ($\sim25~\%$), with the relation derived by \citet{McDermid+2015} from the analysis of the ATLAS$^{3D}$ sample ($z \lesssim0.1$ and $9.5<log(M_{dyn}/M_{\sun})<12$) by means of the PPXF code \citep{Cappellari&Emsellem2004}. 
 At $z\lesssim0.06$, for a given velocity dispersion, our mass-weighted ages are instead older than the ones deduced by \citet{Thomas+2010}. This can be due to the fact that, differently from our approach, they analyzed a sample of $\sim~$3600 morphologically selected galaxies (i.e. MOSES ETGs at $0.05 \leq z \leq 0.06$, see \citealp{Schawinski+2007}), without restriction to the most passive objects and also to the fact that, using the Lick indices, they derive SSP-equivalent instead of mass-weighted ages.\\
It is also interesting to note that our average light-weighted ages for $z\lesssim0.06$ are $\sim~$11.6 Gyr for all of the four mass bins and that these values are in agreement with the ones obtained at similar redshifts by \citet{Conroy+2014}, who applied the stellar population synthesis (SPS) model developed by \citet{Conroy&VanDokkum2012a} to a sample of nearby ($0.025<z<0.06$) SDSS DR7 ETGs, with masses comparable to ours (i.e. $10.70<log(M/M_{\sun})<11.07$).\\

\begin{figure}[!t]
\begin{center}
\includegraphics[width=1\hsize]{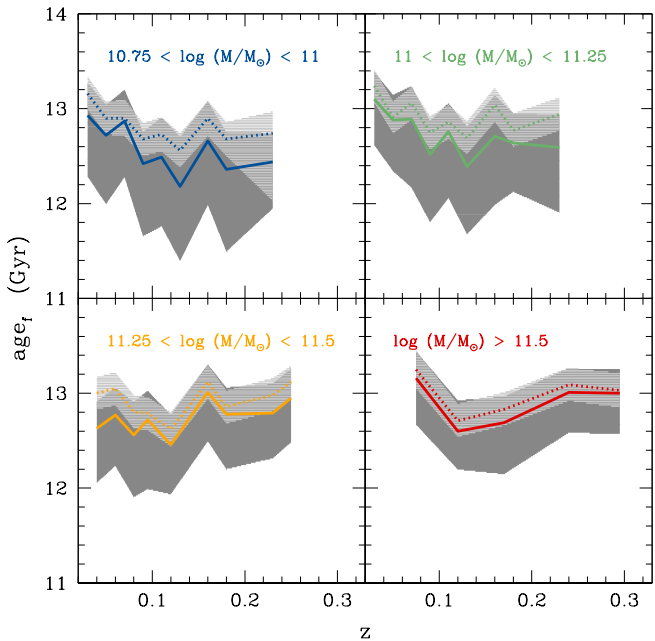}
\caption{$age_{f}$-redshift relations for the four mass bins. Solid and dotted curves (color coded as in Fig. \ref{fig:zage}) are the ages of formation as a function of redshift referring, respectively, to BC03 and MS11 spectral synthesis models. Dark-grey and light-grey shaded regions are the $1\sigma$ dispersions calculated starting from the 16th (P16) and 84th (P84) percentiles of the mass fraction cumulative function for BC03 and MS11 models, respectively.}
\label{fig:ageformation}
\end{center}
\end{figure}

The derived mass-weighted ages imply a very early epoch of formation. A further confirmation of this comes from Fig. \ref{fig:ageformation}, which illustrates the ETGs mass-weighted ages of formation (see eq. \ref{eq:agef}) as a function of mass and redshift (in this case, the term $age_{U}(z)$ in eq. (\ref{eq:agef}) is the age of the Universe at the redshift of each of the available median stacked spectra), together with the $1\sigma$ dispersions calculated from the 16th and 84th percentiles of the mass fraction cumulative distribution. As it is possible to note, the mass-weighted ages of formation are very high, with an increase of $\sim~$0.4 Gyr from low to high masses. 
Taking into account the 1 $\sigma$ dispersion on the SFHs, we also find that the slopes of the $age_{f}-z$ relations are compatible with zero, indicating that, at a fixed mass, the analyzed galaxies have similar $z_{F}$, especially in the highest mass bins. 
However, it is important to note that our main results concerning SFHs do not
rely on the fact that galaxies in the same mass bin have
similar formation epochs,  
since the spectral fits are realized separately on each mass and
redshift bin. \\

Using MS11 models, we find the same trends of the mass-weighted $age_{f}$ with mass and redshift. In particular, in this case $age_{f}$ are only $\sim0.2$ Gyr older than the ones provided by BC03 spectra, on average. In all two cases, the independence of $age_{f}$ of redshift is also an indication that we are not biased towards young galaxy progenitors going to higher redshift, and thus that our sample is not affected by the so-called progenitor bias (\citealp{VanDokkum&Franx1996}).\\

\subsection{Star formation histories}
\label{subsect:sfh}
\begin{figure}[!t]
\begin{center}
\includegraphics[width=1\hsize]{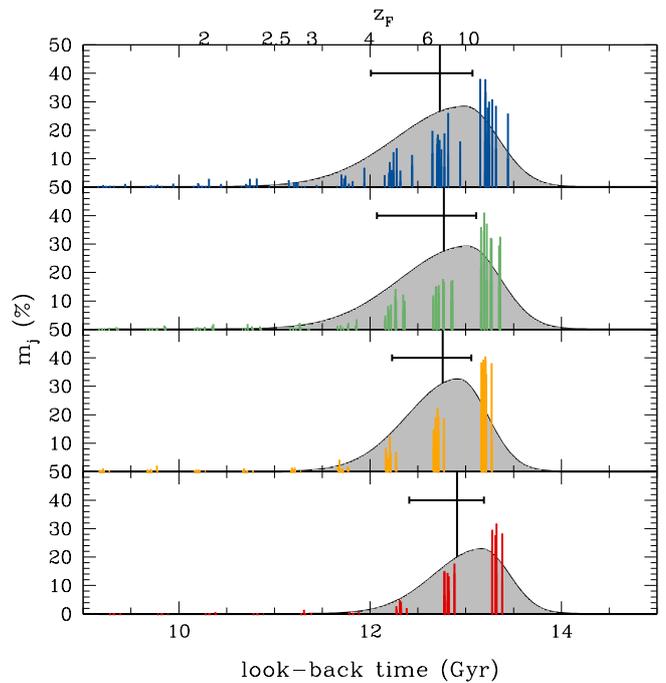}
\caption{Mass fractions $m_{j}$ as a function of look-back time for the four mass bins, with mass increasing from top to bottom (in the case of BC03 models). Colored vertical lines (color-coded as in Fig. \ref{fig:zage}) are the $m_{j}$ obtained from the full-spectrum fitting for each mass bin, put in phase according to their P50. In each mass bin, black vertical lines mark the P50 of the distribution, obtained from data as an average on the P50 of all the redshift bins, while black horizontal lines are the median dispersions [P50 -- P16] and [P84 -- P50]. The corresponding asymmetric gaussians (grey) are overplotted to the $m_{j}$ distributions. On the top of the figure, the $z_{F}$ corresponding to each $age_{f}$ are indicated.}
\label{fig:gauss}
\end{center}
\end{figure}

In this Section we investigate the trend with mass of the SFHs by looking at the mass fractions provided by the full-spectrum fitting ($m_{j}$) as a function of the mass-weighted age of formation, $age_{f}$.
In particular, 
we consider all the redshifts of a given mass bin together, in order to visualize the shape of the SFH as a function of mass. 
The behaviour of the $m_{j}$ distribution is illustrated in Fig. \ref{fig:gauss}, where asymmetric gaussians - constructed starting from the 50th (P50), 16th (P16) and 84th (P84) percentiles of the mass fraction distribution cumulative functions - are overplotted to the $m_{j}$ to better visualize the distribution (P16 and P84 allow to compute the dispersion of the distribution within 1$\sigma$).
Given a mass bin, P50 is the median on the P50 of all the available redshifts, while the asymmetric dispersions are defined as the median differences [P84 -- P50] and [P50 -- P16], averaged in the same way. Table \ref{tab:percentiles} lists the calculated values for the four mass bins. 
To best retrieve the shape of the SFHs and minimize the dependence on possible small differences in $z_{F}$, 
we decide to put the $m_{j}$ distributions in phase, by synchronizing the P50 of each redshift bin to the median P50 of each mass bin, in order not to rely on the assumption that galaxies in a given mass bin have the same formation epochs.\\  
From the figure it is possible to note that the median of the asymmetric gaussians slightly increases with mass and that the width of the gaussians decreases from low to high mass. 
In particular, we find that the width [P84 -- P16] decreases from $\sim~$1 Gyr to $\sim~$0.7 Gyr from low to high masses, while P50 increases by $\sim~$0.2 Gyr. MS11 models produce slightly higher P50, which increase by $\sim~$0.1 Gyr with mass, and dispersions [P84 -- P16] of $\sim0.35$ Gyr, regardless of mass. 

\begin{figure}[!t]
\begin{center}
\includegraphics[width=1\hsize]{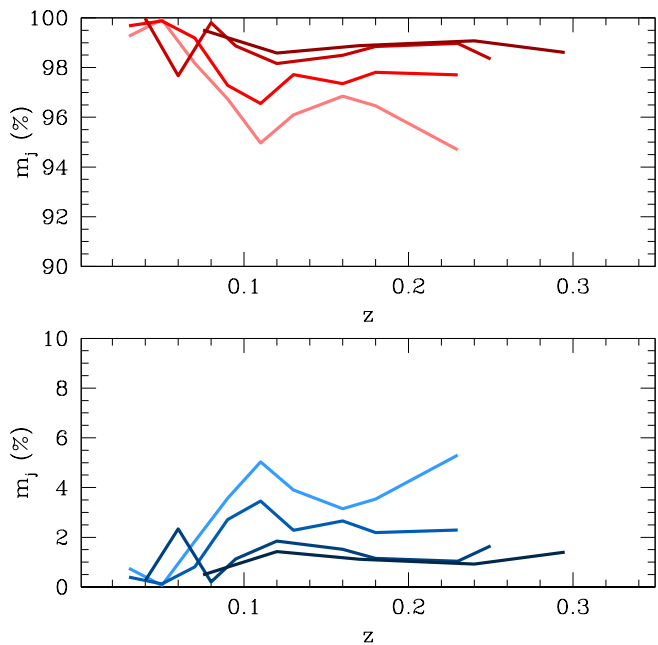}
\caption{Mass fractions $m_{j}$ recovered from the full-spectrum fitting as a function of redshift in the case of BC03 models. Red and blue curves refer, respectively, to stellar populations older and younger than 5 Gyr, with darker colors standing for higher stellar masses.}
\label{fig:mass_fract}
\end{center}
\end{figure}

\begin{figure}[!t]
\begin{center}
\includegraphics[width=1\hsize]{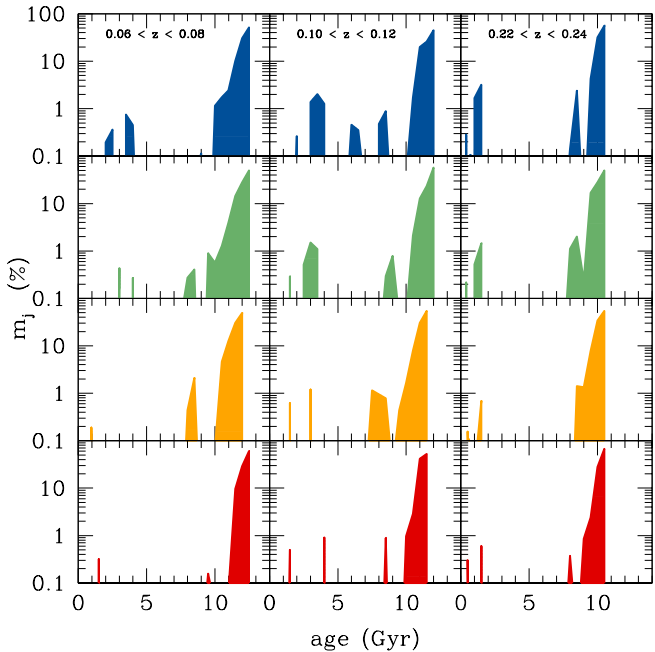}
\caption{SFHs derived from median stacked spectra. Mass increases from top to bottom, redshift increases from left to right, as indicated. Note that no significant episodes of SF occurred after the main SF event. The mass fractions relative to stellar population younger than 5 Gyr is very low (i.e. $\lesssim5~\%$).}
\label{fig:shapesfh1}
\end{center}
\end{figure}

Considering these results and our largest uncertainty on age  $rms_{set}$ (see Table \ref{tab:unc}), the derived trends show that massive galaxies have already formed $\sim~$50$~\%$ of the stellar mass by $z\gtrsim5$ ($z\gtrsim3$ if the $\sim$ 0.5 Gyr systematic is taken into account, see Sect. \ref{stacked_individual}). 
In addition, this percentage of mass is formed slightly earlier in more massive galaxies than in less massive ones, with the former also having shorter star formation histories than the latter (see Fig. \ref{fig:gauss} and Table \ref{tab:percentiles}).\\
We quantify the typical SFR of the analyzed galaxies defining, for each mass bin, the quantity 
\begin{equation} 
\langle SFR \rangle_{68~\%}=0.68\cdot M/\Delta t~(68~\%)~M_{\sun}~yr^{-1}~~, 
\end{equation}
basing on the assumption that the $68~\%$ of stellar mass is formed whithin the $68~\%$ time interval $\Delta t$ around the peak of the SFH. We find that the typical SFRs increase for increasing mass from $\sim50$ to $\sim370$ $M_{\sun}~$yr$^{-1}$ (SFR$~\sim140-750~M_{\sun}~$yr$^{-1}$ for MS11 models).\\
The evolutionary picture emerging from the analysis of the star formation histories can be also deduced from Fig. \ref{fig:mass_fract}, which illustrates the mass fraction relative to the stellar populations older or younger than 5 Gyr. Regardless of redshift and mass, we derive that $\lesssim~$6$~\%$ of the stellar mass in our galaxies come from stellar populations younger than 5 Gyr ($\lesssim~$20$~\%$ when we consider light fractions instead of mass fractions) and that this fraction decreases to $\lesssim4~\%$ and  $\lesssim1~\%$ if we consider a threshold of 1 and 0.5 Gyr, respectively.  Moreover, the old and the young mass fractions are specular to each other, with the young one decreasing with both mass ad cosmic time, contrary to the old one: this suggests that more massive galaxies assembled a higher fraction of their mass earlier than less massive systems. The fact that the analyzed galaxies have not experienced significant bursts of star formation in recent times is also shown in Fig. \ref{fig:shapesfh1} (it is worth noting that these bursts would be recognized by the full-spectrum fitting, as discussed in Sect. \ref{sect:sim}).
MS11 spectral synthesis models confirm these trends, providing a similar low percentage of stellar mass below 5 Gyr (i.e. $<~7~\%$) and an even smaller contribution ($<0.1~\%$) of the stellar populations younger than 1 Gyr to the total mass (which also implies slightly older mass-weighted ages).

 \begin{table*}
\caption{Number of galaxies, statistical, evolutionary and physical properties of the four mass bins of our sample (in the case of BC03 models).}
\renewcommand{\arraystretch}{1.4}
\begin{tabular}{cccccccccc} 
 \hline
$log(M/M_{\sun})$&$\#~$Galaxies &P50 (Gyr)&[P84-P16] (Gyr)&$\tau$ (Gyr)&$\tau$ (Gyr)& $\langle Z\rangle_{mass}$&$A_{V}$ (mag)& $\sigma~$(kms$^{-1}$) \\
&&&&[Expdel]&[Expdelc]&&&&\\
\hline
 $10.75-11$ &12462&12.73&1.06&0.44&0.80&0.028&0.13&208\\
 $11-11.25 $& 8064&12.77&1.04&0.44&0.70&0.029&0.1&235\\
 $11.25 -11.5$& 3004&12.76&0.83&0.40&0.65&0.030&0.05&261\\
 $>11.5$&990&12.91&0.7&0.37&0.60&0.030&0.03&288\\
 \hline
 \end{tabular}
 \label{tab:percentiles}
 \end{table*}

\subsection{The shape of the star formation history} 
\label{subsec:shapesfh}

The SFH of individual galaxies has been often approximated by simple declining exponential functions $SFR(t)\propto exp(-t/\tau)$, where $t$ is the galaxy age and $\tau$ is the star formation timescale (\textit{Expdel} hereafter).
 Despite the success of these models to estimate the stellar masses of nearby spiral galaxies \citep{Bell&DeJong2001}, recently many results have began to highlight their limitations, especially when applied to higher redshift samples (\citealp{Papovich+2001}, \citealp{Shapley+2005}, \citealp{Stark+2009}). Since then, other functional forms have been considered: for example, delayed-$\tau$ (exponentially delayed) models (with $\tau$ being the SF timescale), in which $SFR(t)\propto \tau^{-2}t~exp(-t/\tau)$ (\citealp{Bruzual&Kron1980}, \citealp{Bruzual&Charlot2003}, \citealp{Moustakas+2013}, \citealp{Pacifici+2013}) or 'inverted-$\tau$' models of the form $SFR(t)\propto exp(+t/\tau)$ (\citealp{Maraston+2010}, \citealp{Pforr+2012}) have been suggested to be a better way to represent the SFH of massive galaxies at intermediate or high redshift.\\
In this Section, we test if these two analytical forms are able to describe the SFH of massive and passive, low redshift ETGs. We also introduce the parametric function (\textit{Expdelc} hereafter):
\begin{equation}
SFR(t)=\tau^{-(c+1)}t^{c}~exp(-t/\tau)~~,
\end{equation}
with $c$ being a real number ranging from 0 to 1 with steps of 0.01, which parametrizes how fast the SFR rises at early ages.

\begin{figure}[!t]
\begin{center}
\includegraphics[width=1\hsize]{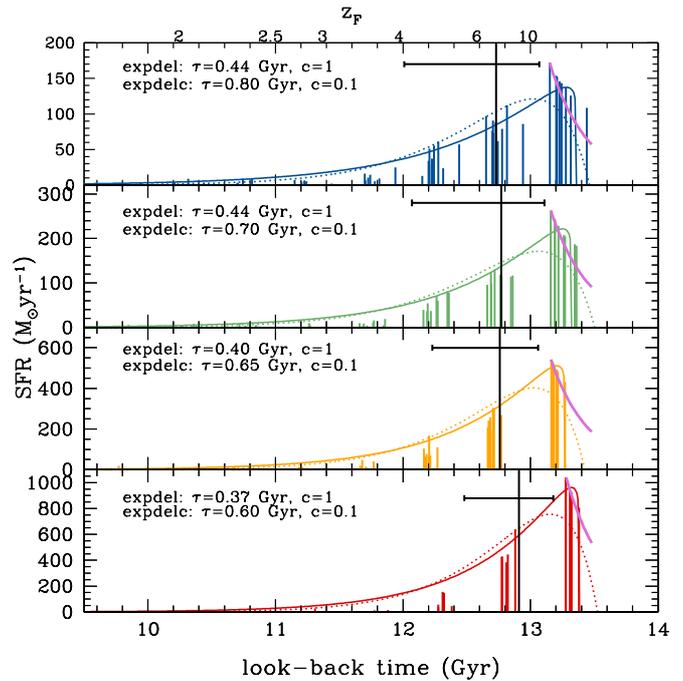}
\caption{SFH for the four mass bins of our sample, with mass increasing from top to bottom (in the case of BC03 models). Colored vertical lines are the SFR ($M_{\sun}~$yr$^{-1}$) derived from the $m_{j}$ provided by the spectral fitting (colors are coded as in Fig. \ref{fig:zage}). In each mass bin, black vertical and horizontal lines are defined as in Fig. \ref{fig:gauss}. In each panel, we show the two best fit models deriving from the assumption of an $Expdel$ (dotted curves) or an $Expdelc$ (solid curves) parametric function to describe the derived SFHs. The best fit parameters are also reported. Pink curves are the inverted-$\tau$ models with $\tau$ = 0.3 Gyr described in the text, extended up to 13.48 Gyr (which corresponds to the age of the Universe in the assumed cosmology).
}
\label{fig:shapesfh}
\end{center}
\end{figure}

The results are shown in Fig. \ref{fig:shapesfh}, in which we illustrate the SFR derived from the mass fractions $m_{j}$ as a function of the look-back time (age of formation), together with the delayed-$\tau$ and the inverted-$\tau$ models.\\ 
For each mass and redshift bin, the observed SFRs are derived using the following equation:
\begin{equation}
SFR_{j}=\frac{m_{j}\times M}{\sum_{j}{m_{j}}\times 0.5 \times10^{9} \times (1-R)}~~ (M_{\sun}~yr^{-1}),
\end{equation}
where $m_{j}$ are the mass fractions obtained from the full-spectrum fitting, $\sum_{j}{m_{j}}$ is the sum of the mass fractions of each mass and redshift bin, $M$ is the total stellar mass, the term $0.5 \times10^{9}~$yr is the time step of our spectral library and R is the 'return fraction', which represents the fraction of mass which is returned to the ISM by supernovae-driven winds and mass losses. In particular, considering both the mass weighted ages of our galaxies (i.e. $\sim~10-13$ Gyr) and the Chabrier IMF assumed in this work, we fix the value of R to 0.5 (see \citealp{Bruzual&Charlot2003}).\\
To perform the data-to-model comparison in the case of exponentially delayed SFHs, we define a grid of models $Expdel(\tau)$ with different SF timescales $\tau$ (going from 0.2 Gyr to 0.7 Gyr  with step of 0.01 Gyr). Then, by minimizing the quadratic differences 
\begin{equation}
\sum_{j}{{[Expdel(\tau, {age_{f}}_{j})-SFR_{j}]}^{2}}~~,
\end{equation}
and
\begin{equation}
\sum_{j}{{[Expdelc(\tau, c, {age_{f}}_{j})-SFR_{j}]}^{2}}~~,
\end{equation}

we find the $\tau$ and $c$ which best reproduce the observed SFH in each mass and redshift bin, by fixing the P50 of data and models. 
Then, as already done in Sect. \ref{subsect:sfh} for the $m_{j}$ distribution, we put the results obtained for all the redshifts in a given mass bin together, also putting their P50 in phase according to the median P50. The median values of $\tau$  and $c$ on all the redshifts are thus taken as the best fit parameters for that given mass bin (see Fig. \ref{fig:shapesfh}). 
We find that the standard delayed-$\tau$ models with shorter $\tau$'s are needed to reproduce the observations for increasing mass. In particular, a value $\tau=0.44$ (with a dispersion of $\pm0.02$ Gyr) is compatible with the SFH of less massive galaxies ($10.75<log(M/M_{\sun})<11.25$), while slightly shorter values (i.e. $\tau=0.37$, with a dispersion of $\pm0.02$ Gyr) should be adopted to match the SFH at higher masses ($log(M/M_{\sun})>11.5$).
The $Expdelc$ parametric form provides a much better fit to the data, with quadratic differences which are smaller even by 2 order of magnitude than in the case of the \textit{Expdel} form. Furthermore, slightly higher values of $\tau$ are found as best fit, decreasing for increasing mass from $0.8 $ Gyr (with a dispersion of $\pm~0.1$ Gyr) to $~0.6$ Gyr (with a dispersion of $\pm~0.1$ Gyr), while the best fit $c$ is $\sim0.1$ (with a dispersion of $\pm~0.05$)  regardless of mass, and reproduces the very fast rise of the SFR (see Table \ref{tab:percentiles}).
The derived values are compatible with the SF timescale $\Delta t~\sim0.4~$Gyr derived by \citet{Thomas+2010} from the measure of the $\alpha$-element abundance\footnote{$[\alpha/Fe]\approx (1/5)-(1/6)\cdot log\Delta t$, see \citet{Thomas+2005}.}. However, it needs to be mentioned that, also using the $[\alpha/Fe]$ ratios, \citet{Conroy+2014} obtained longer SF timescale ($\Delta t \sim~$0.8 Gyr) for similar masses. 
In addition, it is worth noting that, as shown in Sect. \ref{sec:simulation}, the full-spectrum fitting tends to broaden the SFH for old stellar populations in the case of BC03 models (see Fig. \ref{fig:simul1}) and thus the derived SF timescales could be effectively overestimated. \\ However, in agreement with the BC03 analysis, also the best fits with MS11 models confirm the short $e$-folding times, which are $\tau \sim0.25$ Gyr (with a dispersion of $\pm0.0016$ Gyr), regardless of mass ($\tau\sim 0.39$ Gyr -- with a dispersion of $\pm~0.03$ -- and $c\sim0.1$ -- with a dispersion of $\pm~0.02$ -- in the case of $Expdelc$, regardless of mass).\\
Fig. \ref{fig:shapesfh} also illustrates 'inverted-$\tau$' models with $\tau=0.3$ Gyr, normalized at the SFR maximum in each mass bin. The increased difficulty in matching these models with the observations may be linked to our poor sampling of the SFH at very early ages.\\
Fig. \ref{fig:shapesfh} confirms that the SFHs derived from our analysis are smooth and concentrated at high redshift. 
However, we are not able to distinguish whether these smooth SFHs are the result of a rapid collapse at high redshift or derive from coeval mergers.
An important consideration is that the smoothness of the derived SFHs could be due to the fact that they derive from median stacked spectra, which represent the average behaviour of the sample. Therefore, the stochastic SF episodes related to the individual galaxies involved in the stack can be washed out and diluted within the stack. Our results have to be taken with a statistical meaning because are based on the
analysis of stacked spectra. High SNR spectra of individual ETGs are required to better
assess their evolution and star formation histories.


\subsection{Metallicities}
\label{subsect:met}
\begin{figure}[!t]
\begin{center}
\includegraphics[width=1\hsize]{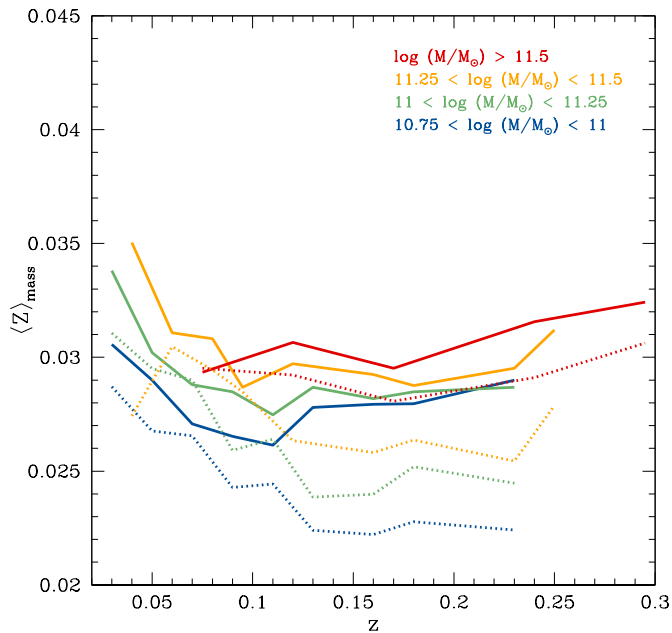}
\caption{Mass weighted metallicities as a function of mass and redshift. Solid and dotted curves refer to BC03 and MS11 models, respectively. Stellar mass increases from blue to red, as indicated in the top right of the figure. }
\label{fig:metz}
\end{center}
\end{figure}

\begin{figure}[!t]
\begin{center}
\includegraphics[width=1\hsize]{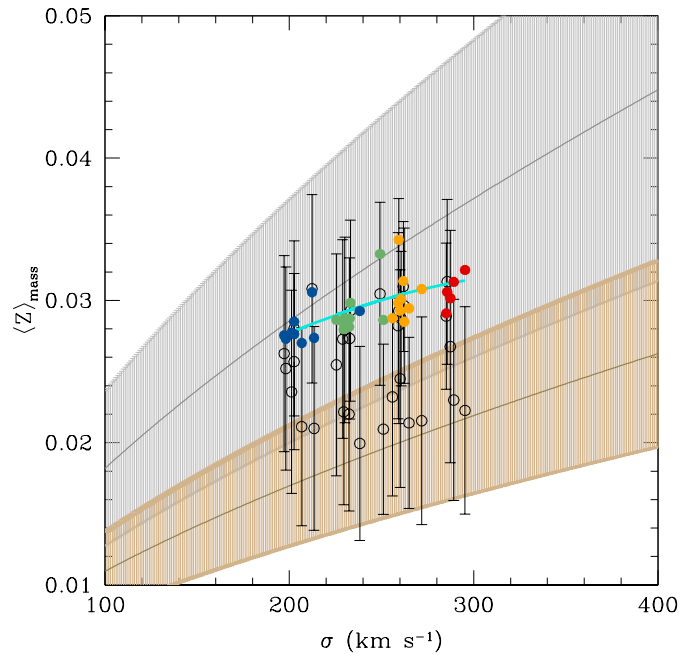}
\caption{Comparison between our mass-weighted metallicities (in the case of BC03 models) and the values reported in the literature. In particular, colored filled circles (color-coded as in Fig. \ref{fig:zage}) are our results (the cyan curve is a second order fit to our data); the grey curve is the scaling relation provided by \citet{Thomas+2010}, with its dispersion (grey shaded region), while black open circles and vertical bars are the measures (and dispersions) performed on our sample by \citet{Gallazzi+2005}. The light-brown curve is the relation derived by \citet{McDermid+2015}, together with its dispersion (light-brown shaded region).}
\label{fig:met_thom}
\end{center}
\end{figure}

From the fit to the median stacked spectra, we also analyze the trends with mass and redshift of the retrieved mass-weighted metallicities $\langle Z\rangle_{mass}$, derived from eq. (\ref{eq:mwz}).
On average, we find supersolar metallicities, with $\langle Z \rangle_{mass}\sim0.029\pm0.0015$, as illustrated in Fig. \ref{fig:metz} (a median value $\langle Z \rangle_{light}\sim~$0.025 is obtained in case of light-weighted metallicities). Our  $\langle Z \rangle_{mass}$ have no clear trend with redshift in a given mass bin, and also do not show a significant dependence on mass (metallicities increase on average by only 0.0025 from the lowest to the highest masses). 
The independence of metallicities of cosmic time is a suggestion that the analyzed galaxies are very old systems, which have depleted all their cold gas reservoir, with no further enrichment of their interstellar medium with new metals (however, we remind that the analyzed interval of cosmic time is not much extended, amounting to $\sim~$3.3 Gyr). The dependence on mass suggests that more massive galaxies are able to retain more metals thanks to their deeper potential wells \citep{Tremonti+2004}. \\We find that MS11 models also provide slightly supersolar metallicities ($\langle Z \rangle_{mass}\sim~$0.027$~\pm~0.0020$, on average), with a more remarkable dependence on mass (metallicity increases by $\sim~$0.005 from the lowest to the highest mass). The three uncertainty estimates $rms_{sim}$, $err_{100}$ and $rms_{set}$ calculated on the mass-weighted metallicities are reported in Table \ref{tab:unc}. \\
In Fig. \ref{fig:met_thom}, instead, we show a comparison between our metallicity estimates and the ones reported in the literature. In particular, we illustrate the scaling relation provided by \citet{Thomas+2010} obtained from their sample of nearby ETGs\footnote{We convert the [Z/H] values given by \citet{Thomas+2010} following  the conversion of \citet{Caputo+2001}: $logZ=[Z/H]-1.7$.}, with its observed average scatter. We also show the median metallicities measured on our sample by \citet{Gallazzi+2005}, together with their MADs. These metallicities are, in all two cases, measured by fitting individual spectral features (Lick indices, \citealp{Burstein+1984}). In particular, \citet{Thomas+2010} derived their values from the measure of 24 Lick indices, while \citet{Gallazzi+2005} used a set of 5 specific Lick indices, including H$\beta$, H${\delta}_{A}$, H${\gamma}_{A}$, $[Mg_{2}Fe]$ and $[MgFe]'$\footnote{Where $[MgFe]'=\sqrt{Mgb~(0.72~Fe5270~+~0.28 Fe5335)}$ and $[Mg_{2}Fe]=0.6~Mg_{2}~+~0.4~log(Fe4531~+~Fe5015)$.}.  \\
It is possible to note that the metallicity estimates are broadly consistent, with a difference $\Delta \langle Z \rangle_{(This~work-Gallazzi)}\sim$+0.0036 and $\Delta \langle Z \rangle_{(This~work-Thomas)}\sim-$0.0032. The figure also illustrates the \citet{McDermid+2015} mass-weighted metallicity -- velocity dispersion relation, which they derive by means of another full-spectrum fitting code (i.e. PPXF), using the Vazdekis synthetic models \citep{Vazdekis+2012}. We find that their metallicities are closer to the solar value (i.e. $Z\sim0.02$), with $\langle \Delta Z\rangle_{(This~work-McDermid)}\sim-$ 0.01.\\We ascribe the discrepancies among these various results mainly to the use of different sets of Lick indices
or synthetic models (in the case of full-spectrum fitting). Therefore, even if there is agreement among the majority of the reported results in predicting supersolar metallicities (i.e. $Z\gtrsim0.025$), which also increase with $\sigma$, it is evident that the absolute values of metallicities depend on the method and the assumptions used for their estimate.

\subsection{Velocity dispersions and dust extinction}
\label{subsect:veldisp}
\begin{figure}[!t]
\begin{center}
\includegraphics[width=1\hsize]{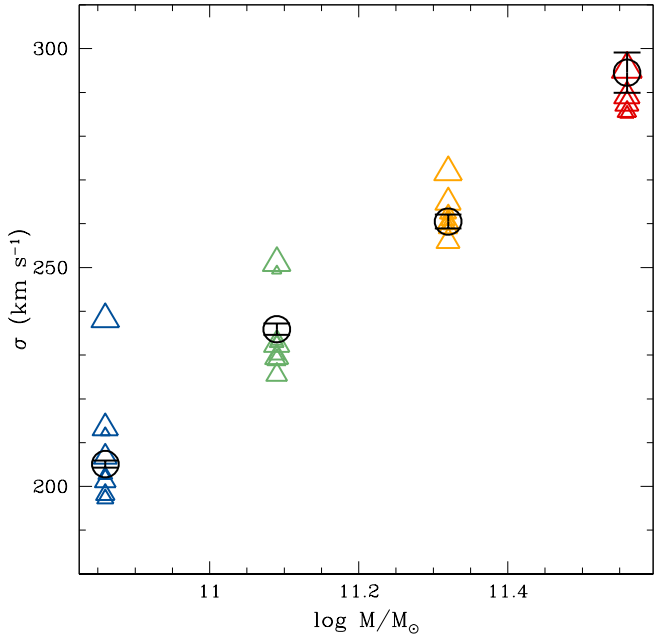}
\caption{Velocity dispersions $\sigma$ as a function of mass (for BC03 models). Colored triangles represent our observed values at different redshift for each mass bin (the values of $\sigma$ are illustrated together with the errors MAD/$\sqrt{N}$), with the size of the triangles increasing for increasing redshift; black open circles are the velocity dispersions derived from the re-reduction of SDSS spectra performed by the Princeton Group, used for comparison.}
\label{fig:vdisp}
\end{center}
\end{figure}

\begin{figure}[!t]
\begin{center}
\includegraphics[width=1\hsize]{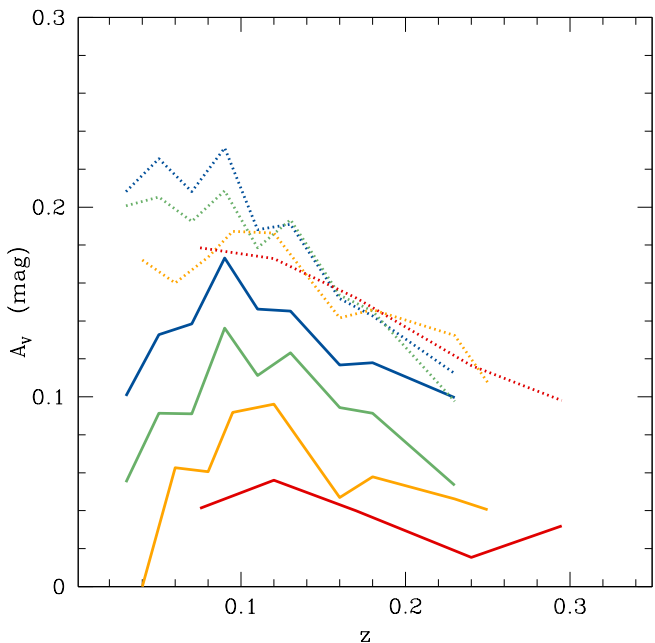}
\caption{Dust extinction $A_{V}$ as a function of redshift and mass. Solid and dotted curves refer to BC03 and MS11 models, respectively. Colors are coded as in Fig. \ref{fig:zage}. }
\label{fig:zav}
\end{center}
\end{figure}

In Fig. \ref{fig:vdisp} we illustrate the trend of the velocity dispersion with stellar mass. In particular, we show the values obtained for all the redshift bins available for a given mass. We find velocity dispersions $\sigma \sim~200-300$ km s$^{-1}$, increasing for increasing  stellar mass. Our results are compared with the median values derived on the same sample by the Princeton group\footnote{More information are available on the website $http://spectro.princeton.edu/$}, who have provided a re-reduction of a subsample of SDSS data. It is possible to note that there is a good agreement between the two estimates. Indeed, the differences $\Delta \sigma_{(This~Work-Princeton)}$ are $\lesssim~$12$~\%$. Furthermore, it is worth noting that the largest discrepancies occur only in correspondence of the extreme low or high redshift bins, which have fewer statistics (i.e. which contain fewer objects), and that MS11 models provide velocity dispersions which are in agreement with the BC03 ones, being only $\sim~$3.5 km s$^{-1}$ lower than them, on average.\\
As it is possible to note in Fig. \ref{fig:zav}, the visual extinction $A_{V}$ obtained from the spectral fit with BC03 models is always $<0.2$ mag and on average $\sim~$0.08 (averaging on the four mass bins), in agreement  with the typical view of early-type galaxies being old and dust-free objects (\citealp{Wise&Silva1996}, \citealp{Saglia+2000}, \citealp{tojeiro+2013}). MS11 models also produce very low $A_{V}$ ($<0.25$ mag) which, however, are higher by $\sim~$0.08 mag than the BC03 ones, on average.
Moreover, in the BC03 case, more massive galaxies have lower $A_{V}$ with respect to less massive ones, while this trend is much less remarkable when MS11 models are used. Although further investigations are needed to confirm the reliability of this trend, it may suggest that more massive galaxies are able to clear their interstellar medium
more efficiently, probably due to more intense feedback processes linked with star formation
or AGN activity.
As usual, the value of $rms_{sim}$, $err_{100}$ and $rms_{set}$ for both the velocity dispersions  and dust extinction are reported in Table \ref{tab:unc}.

\begin{center}
\begin{table*}
{\small
\hfill{}
\caption{Uncertainties on ages, metallicities, velocity dispersions and dust extinction provided by the three different estimates described in the text (in the case of BC03 models).}
\renewcommand{\arraystretch}{1.4}
\begin{tabular}{lccccc} 
 \hline
Uncertainty& $\langle t \rangle_{mass}$ (Gyr) & $Z$ & $\sigma$ (km s$^{-1}$) & $A_{V}$ (mag) \\
\hline
$rms_{sim}$ &$\pm~$0.3& $\pm~$0.0001 &$\pm~$1.3 & $\lesssim10^{-4}$\\
$err_{100}$ & $\pm~$0.05 &$\lesssim~$0.0002& $\pm~$0.3 &$\pm~$0.002\\
$rms_{set}$ &$\pm~$0.6& $\pm~$0.0015& $\pm~$1.75& $\pm~$0.030 \\
\hline
\label{tab:unc}
\end{tabular}}
\end{table*}
\end{center}

\section{Testing the fitting of individual spectra}
\label{stacked_individual}
\begin{figure}[!t]
\begin{center}
\includegraphics[width=1\hsize]{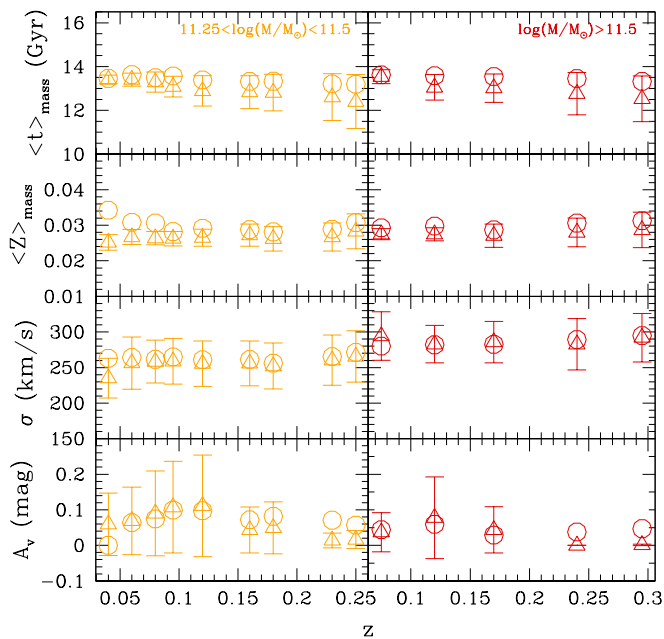}
\caption{Comparison between the median ages, metallicities, velocity dispersions and dust extinction derived from median stacked spectra (open circles) and individual spectra (open triangles) in the case of BC03 models (vertical bars are the MADs on the results from individual spectra) for the two mass bins $11.25<log(M/M_{\sun})<11.5$ (left) and $log(M/M_{\sun})>11.5$ (right).}
\label{fig:stackindiv}
\end{center}
\end{figure}

Since median stacked spectra are very useful to increase the SNR of the observed spectra, in this Section we verify if the results obtained using median stacked spectra are consistent with the ones derived by fitting individual spectra (we performed this check in the case of BC03 models). In particular, we compare the $\langle t\rangle_{mass}$, $\langle Z\rangle_{mass}$, $A_{V}$ and $\sigma$ derived so far through the median stacked spectra, with the median of their distributions obtained studying individual objects.
We restrict our analysis only to the two most massive bins (i.e.  $<11.25<log(M/M_{\sun})<11.5$ and $log(M/M_{\sun})>11.5$), both because they are the most interesting ones to investigate galaxy formation and evolution and for computational feasibility (since they contain fewer objects, see Table \ref{tab:percentiles}). \\ Fig. \ref{fig:stackindiv} illustrates what derived for $\langle t\rangle_{mass}$, $\langle Z\rangle_{mass}$, $\sigma$ and $A_{V}$ at $<11.25<log(M/M_{\sun})<11.5$ and $log(M/M_{\sun})>11.5$.\\ As it is possible to note, in most of the cases the results of median stacked spectra are overlapped to the ones of individual spectra within the dispersion of these latter. 
We found that ages, metallicities and velocity dispersions are higher in the case of median stacked spectra of $\lesssim10~\%$ (the average age differences between individual and stacked spectra are smaller than the uncertainty produced on age by the full-spectrum fitting method, i.e. $\pm~$0.6 Gyr), $\lesssim 15~\%$, $\lesssim 3~\%$ and $\lesssim 15~\%$ respectively.\\ 
Moreover, taking into account the systematic errors derived for the various quantities under analysis (see Table \ref{tab:unc}), we find that the bias between the quantities inferred from median stacked and individual spectra is not significant, with a significance of $\sim0.5~\sigma$ for age, $\sim1.1~\sigma$ for metallicity, $\sim0.05~\sigma$ for dust extinction. In the case of velocity dispersion, it is $\sim2.2~\sigma$  for $11.25<log(M/M_{\sun})<11.5$ and $\sim0.5~\sigma$ for $log(M/M_{\sun})>11.5$. The discrepancy between individual and median stacked spectra is thus small, confirming that our sample is selected to be homogeneus. For this reason, we expect the properties of the median stacked spectra to be similar to ones of each individual spectrum entering the stack.
This test demonstrates that the procedure of stacking spectra do not introduce significant bias on the retrieved evolutionary and physical properties, since the produced shifts mostly lie within the dispersion produced by the fit to individual spectra. Therefore, analyzing stacked spectra is essentially equivalent to analyze individually each spectrum that contributed to the stack.\\
Finally, an important consideration is that, if we consider the small 0.4 -- 0.5 Gyr systematic introduced by stacked spectra, the formation redshifts derived from this analysis decrease to $z\gtrsim3$.

\section{The question of progenitors}
Linking galaxies at $z\sim0$ with their progenitors back in cosmic time is essential 
to derive a coherent evolutionary pattern and understand the formation process of different
galaxy types. In this work, we attempt to place constraints on the progenitors of nearby 
($z \lesssim0.3$), very massive ($log(M/M_{\sun})>10.75$) and passive ETGs 
by exploiting their SFH, reconstructed from the full spectrum 
fitting analysis described in previous Sections. The aim is to verify if the inferred 
properties of the progenitors fit into an evolutionary path consistent with other observational
constraints. In this regard, we first infer the properties of the star forming progenitors 
from which today's massive ETGs originated, and then follow their evolution until they
terminate the star formation activity and become quiescent descendants at later cosmic 
times. \\
As already said, our method is not able to distinguish if the analyzed massive and passive galaxies formed with a monolithic-like collapse at high redshift or if coeval mergers occurred to build their stellar mass. In this regard, it is important to clarify that the results described in the following sections implicitly assume that no coeval merger events have happened during the formation history of the analyzed galaxies. Furthermore, we illustrate the results derived assuming the $Expdelc$ as the parametric form describing the observed SFHs.

\subsection{The star-forming progenitor phase}
\label{subsect:pro}
\begin{figure}[!t]
\begin{center}
\includegraphics[width=1\hsize]{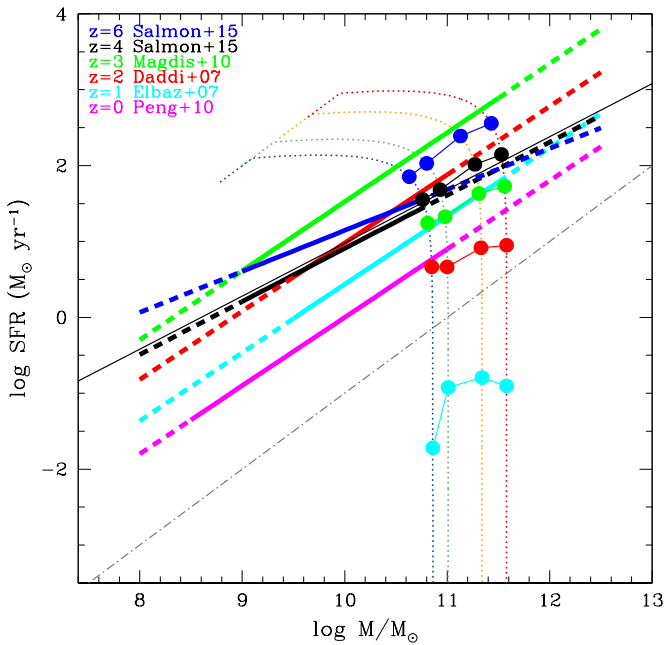}
\caption{Evolving SFR -- mass curves (in the case of BC03 models and for the $Expdelc$ parametric form). The four dotted curves are the evolving SFR -- mass relations for the four mass bins of our sample (deduced from the SFHs of Fig. \ref{fig:shapesfh}) as a function of cosmic time (from left to right), color coded as in Fig. \ref{fig:shapesfh}. Blue, black, green, red, cyan and magenta lines are the SFR -- mass relations deduced by different authors at various redshifts (as indicated in the top left of the figure), within their observed mass ranges (solid lines). Filled circles are the SFR for the four mass bins at various $z$, corresponding to the ones reported in the top left of the figure (note that, at $z\sim0$,  we derive $log(SFR)<-3$, thus the values at this redshift are not included in the plot). 
The grey dashed-dotted line represents the level of SFR at which the galaxies can be considered completely quiescent.}
\label{fig:sfrm}
\end{center}
\end{figure}

Starting from the shapes of the SFHs illustrated in Sect. \ref{subsec:shapesfh} (Fig. \ref{fig:shapesfh}), we display
the inferred evolutionary path of our sample ETGs in the SFR -- stellar mass plane. 
Fig. \ref{fig:sfrm} shows the evolutionary tracks separately for the four mass bins of 
our sample as a function of cosmic time, together with the main sequences (MS) of 
star-forming galaxies available in the literature as a function of redshift. All the 
MS relations have been normalized to the \citet{Chabrier2003} IMF adopted in this work.  

\begin{figure}[!t]
\begin{center}
\includegraphics[width=1\hsize]{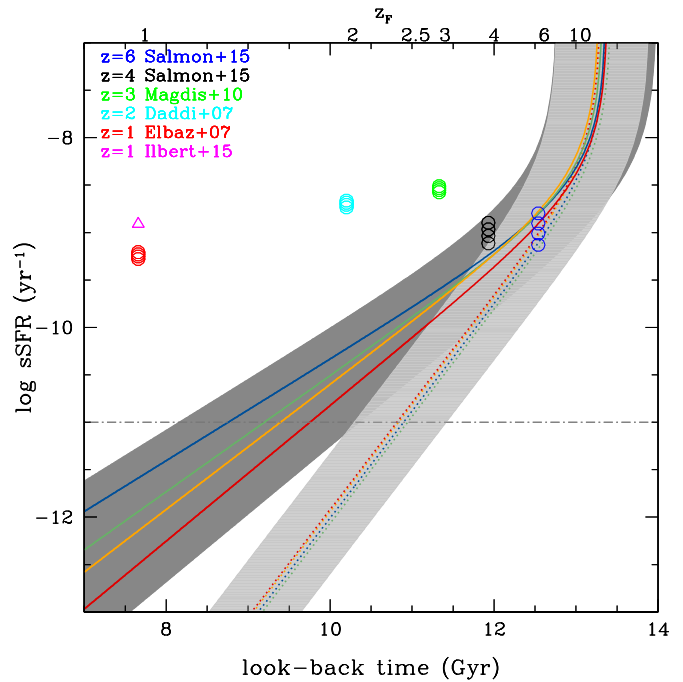}
\caption{Evolving sSFR -- $z$ relations for the $Expdelc$ parametric form. Solid and dotted curves are the sSFR -- $z$ relations for the four mass bins (color coded as in Fig. \ref{fig:shapesfh}) for BC03 and MS11 models, respectively. Blue, black, green, cyan and red open circles (and violet open triangle) are the sSFR estimates obtained by various authors at different redshifts (for the four mass bins), as indicated in the top left of the figure. The grey horizontal dashed-dotted line marks the level of sSFR at which a galaxy is in general considered completely quiescent (sSFR $\lesssim10^{-11}~$yr$^{-1}$). Dark-grey and light-grey shaded regions represent the uncertainty on the look-back time, associated to the results from BC03 and MS11 models, respectively.}
\label{fig:ssfrz}
\end{center}
\end{figure}

Several important results emerge from Fig. \ref{fig:sfrm} (which illustrates the case of BC03 models). The inferred SFRs were high in 
the past, with typical levels $\langle SFR\rangle_{68~\%}$ up to $350-400$ $M_{\sun}~$yr$^{-1}$ for the most massive bin, but broadly consistent with 
those of MS galaxies at the same redshifts and masses. At early epochs ($z\sim3-6$), the SFRs increase with mass and, for a given
redshift, the slope of the SFR--mass correlation is similar to that of the MS. An example
is represented by the comparison with the results of \citet{Salmon+2015}, who used CANDELS 
data to derive the MS relation at $3.5 \lesssim z \lesssim 6$. Our results suggest that
the progenitors of our ETGs were already forming stars at very high rates at $z>3-4$.
It is worth noting that such early formation epochs ($3\lesssim z \lesssim 10$) are also 
derived by \citet{McDermid+2015} for ATLAS$^{3D}$ ETGs with masses comparable to those
of our sample. Fig. \ref{fig:sfrm} also shows that after $\sim400~$Myr, the SFRs of our ETGs 
decrease rather rapidly to a few $M_{\sun}~$yr$^{-1}$ at $z\sim2$, and that this tends
to occur slightly earlier for the galaxies in the highest mass bin (where a deviation from the
slope of the SFR -- mass relation is present). The complete quiescence is achieved at later
cosmic times. In particular, if the SFR at which a galaxy of a given mass is considered quiescent is defined as SFR(M)$~\sim~10^{-11}~$yr$^{-1}~\cdot~M/M_{\sun}$, our sample 
ETGs become inactive at $z\sim1.5-2$.\\ 
These evolutionary trends are apparent in the complementary Fig. \ref{fig:ssfrz}, which illustrates the specific star formation rate (sSFR, i.e. SFR per unit mass) 
for the four mass bins as a function of look-back time, together with the literature values (derived from the SFRs of Fig. \ref{fig:sfrm}). This shows that the peak of star formation
activity of the very massive and passive ETGs occurred at $4\lesssim z \lesssim 7$. It is worth noting that MS11 models provide a similar evolution for the sSFR at $z>6$ (with SFR up to $\sim~700-750~M_{\sun}~$yr$^{-1}$), even if a steepening with respect to the BC03 predictions occurs at lower redshifts, suggesting slightly higher redshifts for the beginning of the quiescent phase 
($z\sim1.5-2$).\\
Either case, we find that the studied galaxies have already formed $\gtrsim75~\%$ (i.e. $M\gtrsim
10^{10.62}M_{\sun}$) of the total stellar mass by $z\sim5^{+3.8}_{-1.4}$, thus we expect their progenitors at this redshift to be also very massive -- in the hypothesis of no coeval mergers.
If these systems are dusty and follow the same star formation scaling relations present
at lower redshifts, the inferred SFRs imply infrared luminosities in the range of the so 
called luminous to ultra-luminous infrared galaxies ($11 \lesssim log(L_{\rm IR}/L_{\sun}) 
\lesssim 13$).

Clearly, the question is whether galaxies with very high SFRs and substantial stellar masses
have been observed at such high redshifts (if no mergers occurr). The answer is probably yes, although the 
information is still limited and fragmentary. Examples are represented by high redshift
millimeter/submillimeter galaxies at $z\sim4-5$ (\citealp{Schinnerer+2008}; \citealp{Daddi+2009}; 
\citealp{Hodge+2015}) and massive starbursts at $z\sim5-6.3$ (\citealp{Walter+2012Nat}; \citealp{Riechers+2013Nat}; \citeyear{Riechers+2015}). 
Also the existence of galaxies at $z>3$ with simultaneous substantial stellar masses and high 
SFRs is suggestive of systems that could evolve rapidly towards a quiescence phase (\citealp{Bethermin+2015}; see also \citealp{Pozzi+2015}). In this regards, also the massive and mostly quenched galaxies 
at $z\sim3$ discovered by \citet{Taniguchi+2015} could fit well into a scenario of earlier star 
formation as inferred from the SFHs of our sample of ETGs. Last but not least, the mere existence of 
QSOs at $z\sim6$ with massive and evolved host galaxies (e.g. \citealp{Fan+2003}, \citealp{Dietrich+2003}, \citealp{Freudling+2003}, \citealp{Goto+2009}, \citealp{Banados+2014}, \citealp{Wu+2015}) may be consistent with this scenario.


\subsection{The quiescent descendent phase}
According to the inferred SFHs (Fig. \ref{fig:shapesfh}), the progenitors of our sample ETGs are expected to be already quiescent by $z\sim1.5-2$ ($z\sim2.5-3$ in the case of MS11 models), with this phase lasting 
without other major episodes of star formation at later cosmic epochs. This implies that 
a population of massive and passive galaxies ($M\gtrsim10^{10.75} M_{\sun}$) 
should already be in place at $z\sim2-4$. 

Such kind of galaxies have been indeed spectroscopically identified up to $z\sim3$ (\citealp{Dunlop+1996},  \citealp{Cimatti+2004Nat}, \citealp{McCarthy+2004}, \citealp{Kriek+2006}, \citealp{Gobat+2012}), and 
subsequently studied in more detail (e.g. \citealp{Cimatti+2008}, \citealp{Saracco+2009}, \citealp{Onodera+2015}, \citealp{Belli+2015}, \citealp{Lonoce+2015}, see also \citealp{Newman+2015sub}). Their properties are broadly consistent with those 
expected in the evolutionary scenario inferred in this work. As a matter of fact, these galaxies 
are massive ($M\sim10^{10.5-11} M_{\sun}$), dominated by old stars ($\gtrsim~$1 Gyr), 
passive or weakly star-forming, with low sSFR, and characterized by spheroidal morphologies and 
surface brightness profiles typical of early-type galaxies. However, at fixed mass, their sizes are, on average, 
smaller than in present-day ETGs, implying a substantially higher stellar mass density (\citealp{Daddi+2005}, \citealp{Cimatti+2008}, \citeyear{Cimatti+2012}, \citealp{Trujillo+2011}).

The archaeological results of our work imply that if the complete quiescence is achieved by $z\sim2^{+0.5}_{-0.3}$ ($z\sim3^{+1}_{-0.6}$ for MS11 models) on average,
nearly quiescent galaxies (sSFR $\sim10^{-11}~$ yr$^{-1}$, up to $10^{-10}~$ yr$^{-1}$) should be 
present also at higher redshifts. This possibility is now strongly supported by the detection of
several massive quiescent galaxy candidates at $3<z<7$ (\citealp{Mobasher+2005}, \citealp{Rodighiero+2007}, \citealp{Wiklind+2008}, \citealp{Mancini+2009}, \citealp{Fontana+2009}, \citealp{Brammer+2011}, \citealp{Muzzin+2013}, \citealp{Straatman+2014}, \citealp{Caputi+2015}, \citealp{Marsan+2015}). Although these galaxies
are too faint for the current spectroscopic sensitivity, their photometric SEDs suggest that
they are massive (up to $M\sim10^{11} M_{\sun}$), old (often with ages close to the
age of the Universe at their redshifts) and have low sSFRs. We note that the existence
of massive galaxies at high redshifts is also a crucial test for the structure formation in the
$\Lambda$CDM cosmological context (\citealp{Steinhardt+2015}).

\subsection{The size of the progenitors}
\begin{figure}[!t]
\begin{center}
\includegraphics[width=1\hsize]{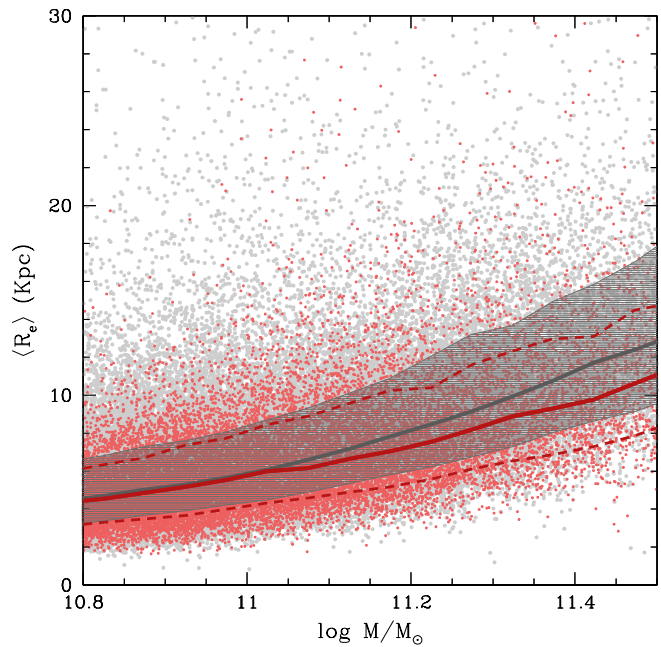}
\caption{Size -- mass relation derived for our sample of massive, passive ETGs (red) and for the parent sample of local galaxies with $log(M/M_{\sun})>10.75$ (grey) (\citealp{Kauffmann+2003} stellar masses are rescaled to the M11 ones). Colored points represent the two samples; solid curves are the P50 of the $R_{e}$ distribution at each mass. The dashed red curves and the shaded grey region includes the 68$~\%$ of the $R_{e}$ distribution of the passive and the parent sample, respectively.}
\label{fig:reff}
\end{center}
\end{figure}

More information comes from the the
sizes of our passive ETGs. Fig. \ref{fig:reff} shows a comparison between 
the effective radii $R_{e}$ of the analyzed galaxies and the ones of the parent population, which we define as the ensamble of SDSS DR4 ETGs with $log(M/M_{\sun})>10.75$ \citep{Kauffmann+2003}\footnote{We rescaled the stellar masses of the parent population, estimated by \citet{Kauffmann+2003}, to M11 stellar masses, by subtracting 0.2 dex to the former.}. In particular, $R_{e}$ 
are taken from the NYU Value-Added Galaxy catalogue \citep{Blanton+2005b}. Moreover, since the majority of 
our passive, massive ETGs are bulge-dominated systems (see Sect. \ref{sec:sample}), we restrict this analysis only to galaxies with this kind of morphology, both in the passive and the parent sample. \\
Fig. \ref{fig:reff} shows that the median $R_{e}$ increases for increasing mass (from 
$\sim~$5 Kpc to $\sim~$ 20 Kpc) and, given a stellar mass, our passive ETGs have median 
sizes smaller than the ones of the parent sample even by $\sim15~\%$ (at the highest masses). Furthermore, also the entire $R_{e}$ distribution is extended to smaller radii in 
the case of passive ETGs, especially for $log(M/M_{\sun})\gtrsim11.5$. 
Reminding that small differences in galaxy size imply large differences in stellar mass 
density,  the derived trends suggest that the ETGs analyzed in 
this work should have formed from higher density progenitors which, in the hypothesis of no coeval mergers, do not increase their mass during the evolution. On the other hand, we cannot exclude that galaxies in the parent sample have experienced more dry mergers than our massive and passive galaxies, which have increased their size across cosmic time (\citealp{Naab+2009}, \citealp{Johansson+2012}). The observed trend is also in agreement with some literature studies at low redshift 
which also suggest that more compact galaxies contain older stellar populations than larger 
ones (e.g. \citealp{Saracco+2009}, \citealp{Shankar&Bernardi2009}, \citealp{Williams+2010}, \citealp{Poggianti+2013},
\citealp{McDermid+2015}). In general, this is qualitatively consistent with the high 
$z_{F}$ inferred by our analysis ($z>5$) since, according to the cosmological evolution
of the baryonic matter density, gas was denser at these cosmic epochs. These findings also agree with recent observations
of compact systems (with various level of SF) at $z\sim2-3$, which could be identified as local massive ETGs progenitors (e.g. \citealp{Daddi+2004}, \citealp{Cattaneo+2013}, \citealp{Finkelstein+2013Nat}, 
\citealp{Marchesini+2014}, \citealp{Nelson+2014Nat}, \citealp{Williams+2015}).
Moreover, recent models of elliptical 
galaxy formation also predict the formation of compact and dense progenitors at high redshifts
(e.g. \citealp{Johansson+2012}, \citealp{Naab+2014} and references therein).

\subsection{The number density of progenitors}

The number density ($\rho_{n}$) of our ETGs can be used as another consistency check in the comparison
between the low redshift galaxy population and their progenitors at earlier epochs.
To do that, we first derived the mass function (MF) of the analyzed galaxies at $z\sim0$ by rescaling the local \citet{Baldry+2008} MF\footnote{We rescaled \citet{Baldry+2008} stellar masses (which are derived from an average on four 
stellar mass estimates, which include the \citet{Kauffmann+2003} ones) to M11 ones.} (of all galaxy types) to the percentage of galaxies of each mass bin towards the parent population (which increases from $\sim23\%$ to $\sim70 \%$ from the lowest to the highest masses
), which is defined as in the previous section, but now including galaxies of any morphology. Then, we evolve this local MF back in cosmic time following the decrease of the stellar mass predicted by the derived SFHs, inferring, at each cosmic time, the number density of galaxies above $log(M/M_{\sun})=10.75$. We verified that the mass completeness of our subsample 
of massive and passive galaxies is consistent with the one of the global sample 
of galaxies above $log(M/M_{\sun})=10.75$ and at $z\lesssim0.3$, on which the $V_{max}$ correction 
was already applied (see \citealp{Baldry+2008}), and that our mass subsamples show
no significant mass evolution as a function of redshift.

\begin{figure}[!t]
\begin{center}
\includegraphics[width=1\hsize]{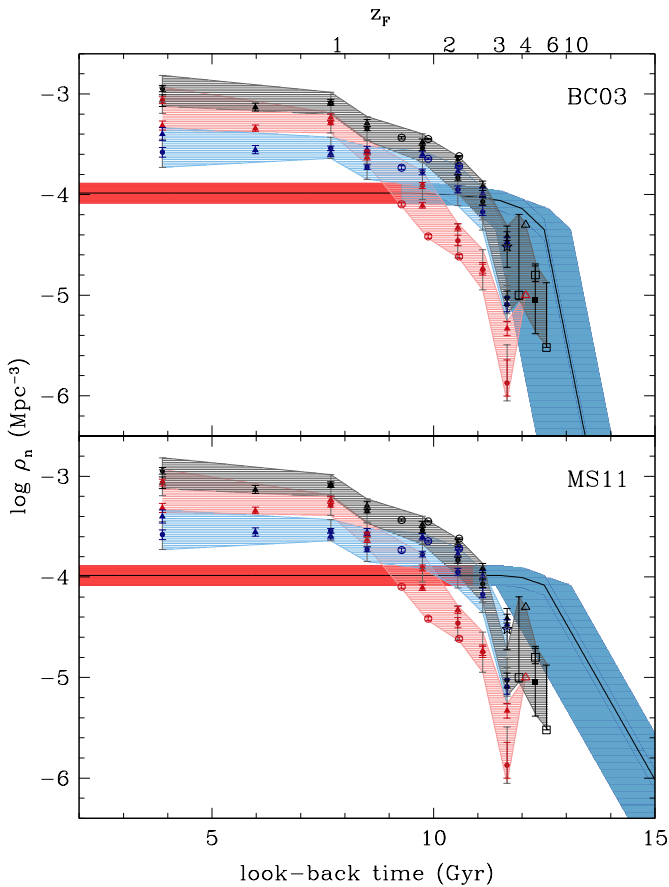}
\caption{Number density above $log(M/M_{\sun})=10.75$. The central colored bands are the $\rho_{n}$ of the ETG progenitors as a function of the look-back time in the case of BC03 (upper panel) and MS11 (lower panel) models, derived from the SFHs of the analyzed ETGs, together with its uncertainty ($\pm~0.1~$dex) and the uncertainty on the look-back time (i.e. $^{+0.6}_{-1.1}~$Gyr, considering the small $\sim0.5$ Gyr systematic introduced by median stacked spectra, see Sect. \ref{stacked_individual}). The red and the blue parts cover, respectively, the redshift interval within which ETGs are completely quiescent or star forming. At each cosmic epoch, the literature results are illustrated with red, blue and black symbols, which refer to the $\rho_{n}$ of the quiescent, the star forming and the global galaxy population (i.e. Star forming$~+~$quiescent galaxies), respectively (all above the same mass threshold). In more detail, closed triangles and cirlces are the \citet{Ilbert+2013} and the  \citet{Muzzin+2013} $\rho_{n}$, respectively; open triangles and circles are the \citet{Mancini+2009} and the \citet{DominguezSanchez+2011} values, respectively; open square and star are the \citet{Grazian+2015} and the \citet{Caputi+2015} values, respectively. The shaded colored regions emphasize the uncertainties on the literature number densities (at each cosmic time, we consider the outermost envelope which includes all the literature estimates, taking into account their uncertainties). It is important to note that the literature $\rho_{n}$ and their associated errors are derived by integrating the GSMF inferred through the $1/V_{max}$ method, and by computing the quadrature sum of its errors. The \citet{Muzzin+2013} error estimates (obtained from the best fit Schechter functions) are also shown in grey.}
\label{fig:ND}
\end{center}
\end{figure}

At $z\sim0$, we obtain $log(\rho_n)\sim-3.98\pm0.1~$Mpc$^{-3}$. \\ Fig. \ref{fig:ND} shows a comparison between the $\rho_{n}$ obtained from this procedure and the literature\footnote{The literature MF have been all rescaled to a \citet{Chabrier2003} IMF and to BC03 stellar masses (M11 stellar masses rely on these spectral synthesis models).}, as a function of both the look-back time and the spectral synthesis models used for the fit (BC03 and MS11 models). In the case of BC03 models we find that, when the analyzed ETGs are completely quiescent and have already formed the totality of their stellar mass ($0\lesssim z\lesssim2$), the $\rho_{n}$ inferred by us is independent of look-back time and is compatible with the literature, not exceeding the total number of passive galaxies per unit volume observed above the same mass threshold. Moreover, the redshift at which we  have no more quiescent galaxies (since they are turning into star forming systems, $z\sim2$) is consistent with the one at which the $\rho_{n}$ of quiescent galaxies observed in literature starts to rapidly decline. At earlier epochs, when the ETGs are still forming stars, the $\rho_{n}$ of the SF progenitors starts to turn down, as the stellar mass formed by the ETGs decreases, and its value at $2\lesssim z\lesssim3$ (i.e. $log(\rho_n)\sim-4.1$ Mpc$^{-3}$) is in agreement with the literature observations at the same redshifts. Going towards the beginning of the SF ($z>3$), $\rho_{n}$ rapidly decreases and, within the uncertainty on the look-back time ($\pm~$0.6 Gyr, see Table \ref{tab:unc}), this decline is in agreement with the one observed in the literature for the global population of SF and quiescent galaxies. \\In the case of MS11 models, the drop of $\rho_{n}$ occurs at slightly earlier times due to the shorter derived SFH.  However the results are, also in this case, in agreement with the low redshift observations, and a marginal consistency with the literature is still present at high $z$.\\

Finally, for each mass bin, we also calculate the star formation rate density (SFRD) of the SF progenitors at $z\sim5$ by multiplying the typical star formation rates $\langle SFR\rangle_{68~\%}$ defined in Sect. \ref{subsect:pro} by the $\rho_{n}$ of the considered mass bin. From this, we infer a global SFRD $log($SFRD$)$ $\sim-1.71~M_{\sun}~$yr$^{-1}~$Mpc$^{-3}$, which is totally consistent, within the uncertainties, to the \citet{Madau&Dickinson2014} SFRD at the same redshift, which amounts to $log($SFRD$)$ $\sim-1.72\pm0.27~M_{\sun}~$yr$^{-1}~$Mpc$^{-3}$ (the uncertainties on the \citet{Madau&Dickinson2014} SFRD are taken from \citealp{Behroozi+2013}). It is however worth noting that, in the case of MS11 models, a higher SFRD, i.e. $log($SFRD$)$ $\sim-1.22~M_{\sun}~$yr$^{-1}~$Mpc$^{-3}$, is derived at the same redshift.

\section{Summary and conclusions}

In this work we analysed the median stacked spectra of a sample of 24488 SDSS DR4 ETGs at $z<0.3$ 
in order to derive the properties of their stellar populations and to reconstruct their
star formation histories. This study focused on the extreme cases of passive ETGs with the
highest stellar masses ($log(M/M_{\sun})>10.75$) which, in turn, represent the
passive envelope of the galaxy population at $0<z<0.3$. The sample was divided into four bins with
increasing stellar mass. The stacked optical spectra were analysed by 
means of the full-spectrum fitting technique using the public code STARLIGHT in order to 
derive constraints in a way complementary to the traditional method of Lick indices based
on a few absorption lines. Using this archaeological approach, we also inferred the properties
of the progenitors of massive and passive ETGs. Our main results can be summarized as follows. 

\begin{itemize}

\item First of all, STARLIGHT was tested against stellar population synthesis models in order 
to assess its reliability in the case of spectra similar to those of ETGs. It was found that 
the software retrieves the stellar population main properties (age, metallicity, SFH, dust
extinction) and the velocity dispersion with a percentage accuracy higher 
than 10$~\%$ for SNR $~\gtrsim~$10 -- 20, even if more complex SFHs are considered. In order to minimize the uncertainties, the STARLIGHT
analysis was applied to stacked spectra with typical SNR$~\sim~$80.\\

\item Mass-weighted stellar ages are very old, increasing with cosmic time from $\sim~$10 to $\sim~$13 Gyr, and show
a clear tendency to increase with mass despite the rather limited mass leverage of our sample,
which is selected to include only the most massive systems ($log(M/M_{\sun})>10.75$). 
This result provides an additional support to the downsizing evolutionary scenario where more massive galaxies are older than less massive ones. The derived ages are broadly compatible with those found with different full-spectrum 
fitting codes (e.g. \citealp{McDermid+2015} and \citealp{Conroy+2014}) and with the Lick indices 
approach (e.g. \citealp{Thomas+2010}; \citealp{Graves+2008}).
The bottom line is that the most massive and passive ETGs represent the oldest galaxies in the present-day
Universe, with ages close to the age of the Universe itself in the most extreme cases.\\
\item Mass-weighted metallicities are slightly supersolar, with a median $Z\sim~0.029\pm0.0015$ ($Z\sim~0.027\pm0.0020$ for MS11 models), increase with stellar mass, and do not show any significant trends with redshift. 
This supports the interpretation that the analyzed galaxies are very old objects which formed 
the bulk of their stars much earlier and did not enrich significantly their interstellar medium 
with new metals at later epochs (however, we remind that we have a limited leverage in cosmic time, i.e. $\sim3.3$ Gyr). This contrasts with the metallicity evolution of star-forming 
galaxies (e.g. \citealp{Maiolino+2008}, \citealp{Mannucci+2009}, \citealp{Foster+2012}, \citealp{Zahid+2013}, \citealp{DeRossi+2015}). Although our metallicities are broadly consistent with 
other results at a fixed stellar mass, a large scatter is present amongst the estimates of the 
metallicity ($Z$) obtained with different methods. \\
\item The SFHs inferred with the full-spectrum fitting suggest that the star formation occurred
during an extended period of time. 
The SFHs are globally compatible with a parametric function of the form $SFR(t)\propto \tau^{-(c+1)}t^{c}~exp(-t/\tau)$, where the typical value of $\tau$ and $c$ are always short, with $\tau$ decreasing from $0.8$ to $0.6$ Gyr (with a dispersion of $\pm~0.1$) from lower to higher masses  and $c\sim0.1$ (with a dispersion of $\pm~0.05$) regardless of mass, reproducing the fast rise of the SFR at the beginning of the SFH. 
Other works highlighted a stronger dependence of the star formation timescale on the mass
(e.g. \citealp{Thomas+2010}, \citealp{McDermid+2015}). \\
\item Based on the inferred SFHs, we derive that the ETGs of our sample formed about 50$~\%$ of 
their stellar mass at early epochs, i.e., on average, at $z\gtrsim5$ (which decrease to $z\gtrsim3$ if the small $0.5$ Gyr systematic introduced by median stacked spectra is taken into account). 
Moreover, the most massive galaxies 
formed it $\sim~$0.2 Gyr ($\sim~$0.1 Gyr for MS11 models) before than less massive systems.\\
\item Low dust extinction ($A_{V}\lesssim0.2-0.25$ mag) are required to fit the spectra. 
In addition, in the case of BC03 models, a trend is present showing that $A_V$ tends
to decrease for increasing stellar mass. The reliability, the significance and the interpretation of this
trend requires further analysis beyond the scope of this paper, but a qualitative
interpretation may be that the most massive galaxies were able to "clear" their interstellar medium
more efficiently probably due to more intense feedback processes due to star formation
and/or AGN activity.\\
\item The stellar velocity dispersions $200 \lesssim \sigma \lesssim 300$ km s$^{-1}$ estimated
with STARLIGHT are consistent with estimates by other groups within the uncertainties, and
increase with stellar mass. \\
\item Based on the SFHs, we reconstructed the mass assembly history and the properties
of the progenitors of the most massive ETGs of the present-day Universe, assuming that no coeval mergers have occurred during the evolution of the analyzed galaxies. The SFHs 
imply that these galaxies were vigorously forming stars and assembled large stellar
masses ($\gtrsim75~\%$ of the total stellar mass) by $z\sim5^{+3.8}_{-1.4}$, with SFR up to $50-370~
M_{\sun}~$yr$^{-1}$ (SFR$~\sim~140-750~M_{\sun}~$yr$^{-1}$ for MS11 models). Possible star-forming progenitors with these characteristics 
have indeed been found in samples of galaxies at $4<z<6.5$ selected in the submm/mm.
The inferred SFHs also predict the existence of quiescent galaxies at $2<z<4$
characterized by large stellar masses and low specific star formation rate. Galaxies 
with these properties have been spectroscopically identified at $z\sim2-3$, and
photometric candidates at higher redshifts also exist.\\
\item The $R_{e}$ of our sample ETGs are significantly smaller than those of the parent sample
of SDSS DR4 ETGs with $log(M/M_{\sun})>10.75$. This implies higher stellar mass 
densities and suggest that they should have formed from high density progenitors\\
\item Based on the number density of the analyzed galaxies in the present-day 
Universe ($log(\rho_n)\sim-3.98~\pm~0.1~$Mpc$^{-3}$), the inferred number densities of the progenitors 
are consistent with the literature, within the uncertainties. Moreover, the mean star formation rate density implied by the 
star-forming progenitors does not violate the \citet{Madau&Dickinson2014} relation at high redshift ($z\sim5$), in the case of BC03 models.\\
\item We find a good agreement among our 
results concerning ages, metallicities and SFHs and the ones obtained using the fit to 
individual spectral features (Lick indices) and $\alpha$-elements abundances. We thus 
suggest the full-spectrum fitting to be a complementary and valid approach to derive
the stellar population properties and the star formation histories of early-type galaxies.

\end{itemize}

ACKOWLEDGEMENTS
{The authors are grateful to Emanuele Daddi
for useful discussion, Maria Cebri\'{a}n for providing the effective radii from the NYU-VAGC 
catalogue and Helena Dom\'{i}nguez-S\'{a}nchez, Kenneth Duncan, Andrea Grazian, Chiara Mancini and Adam Muzzin for providing their mass functions. AC is also gratetful to Alfonso Veropalumbo for helpful discussion.
We also acknowledge the support of the grant PRIN MIUR 2010 {\em The 
dark Universe and the cosmic evolution of baryons: from current surveys to Euclid.}
Funding for the SDSS and SDSS-II has been provided by the Alfred P. Sloan Foundation, the Participating Institutions, the National Science Foundation, the U.S. Department of Energy, the National Aeronautics and Space Administration, the Japanese Monbukagakusho, the Max Planck Society, and the Higher Education Funding Council for England. The SDSS Web Site is http://www.sdss.org/. The SDSS is managed by the Astrophysical Research Consortium for the Participating Institutions. The Participating Institutions are the American Museum of Natural History, Astrophysical Institute Potsdam, University of Basel, University of Cambridge, Case Western Reserve University, University of Chicago, Drexel University, Fermilab, the Institute for Advanced Study, the Japan Participation Group, Johns Hopkins University, the Joint Institute for Nuclear Astrophysics, the Kavli Institute for Particle Astrophysics and Cosmology, the Korean Scientist Group, the Chinese Academy of Sciences (LAMOST), Los Alamos National Laboratory, the Max-Planck-Institute for Astronomy (MPIA), the Max-Planck-Institute for Astrophysics (MPA), New Mexico State University, Ohio State University, University of Pittsburgh, University of Portsmouth, Princeton University, the United States Naval Observatory, and the University of Washington.}

%


\bibliographystyle{aa}
\bibliography{art_starlight}
\begin{appendix} 

\end{appendix}

\end{document}